\documentclass[fleqn,usenatbib]{mnras}

\usepackage{newtxtext,newtxmath}

\usepackage[T1]{fontenc}

\DeclareRobustCommand{\VAN}[3]{#2}
\let\VANthebibliography\thebibliography
\def\thebibliography{\DeclareRobustCommand{\VAN}[3]{##3}\VANthebibliography}

\usepackage{graphicx}	
\usepackage{amsmath}	

\usepackage{tikz,xparse}
\NewDocumentCommand{\linkterms}{ O{1ex} m O{} m m }{
  \tikz[remember picture,baseline=(A.base)]{\node[inner xsep=0pt] (A) {$#2$};}
  #3
  \tikz[remember picture,baseline=(C.base)]{\node[inner xsep=0pt] (C) {$#4$};}
  \tikz[remember picture,overlay]{
    \draw (A.south) -- ([yshift=-#1]A.south) -- coordinate (Z) ([yshift=-#1]C.south) -- (C.south);
    \draw (Z) -- +(0,-#1) node[below] {$#5$};
  }
}

\usepackage{booktabs}
\usepackage{array,tabulary,tabularx,multirow}
\usepackage{tikz,pgfplots}
\pgfplotsset{compat=1.18}
\usepackage{epstopdf}
\usepackage{graphicx}
\usepackage{subcaption}
\usepackage{anyfontsize}

\newcommand*{\mlinei}[1]{%
\begingroup
    \renewcommand*{\arraystretch}{1.1}%
   \begin{tabular}[c]{@{}>{\centering\arraybackslash}p{4.5cm}@{}}#1\end{tabular}%
  \endgroup
}

\newcommand*{\mlineii}[1]{%
\begingroup
    \renewcommand*{\arraystretch}{1.1}%
   \begin{tabular}[c]{@{}>{\centering\arraybackslash}p{2cm}@{}}#1\end{tabular}%
  \endgroup
}

\title[Array Mutual Coupling Effects in 21-cm Experiments]{Uncovering the Effects of Array Mutual Coupling in 21-cm Experiments with the SKA-Low Radio Telescope}

\author[O'Hara et al.]{\parbox{\textwidth}{
Oscar S.D. O'Hara$^{1,2}$,\thanks{E-mail: osdo2@cam.ac.uk}
Quentin Gueuning$^{1,2}$,\thanks{E-mail: qdg20@cam.ac.uk}
Eloy de Lera Acedo$^{1,2}$,
Fred Dulwich$^{1}$,
John Cumner$^{1,2}$,\newline
Dominic Anstey$^{1, 2}$,
Anthony Brown$^{3}$, 
Anastasia Fialkov$^{2,4}$,
Jiten Dhandha$^{2,4}$,
Andrew Faulkner$^{1}$,\newline
Yuchen Liu$^{1,2}$
\\
\em{\small\vspace*{-0.5\baselineskip}
$^{1}$Cavendish Astrophysics, University of Cambridge, Cambridge, CB3 0HE, UK \\ \vspace*{-0.5\baselineskip}
$^{2}$Kavli Institute for Cosmology in Cambridge, University of Cambridge, Cambridge, CB3 0HA, UK\\ \vspace*{-0.5\baselineskip}
$^{3}$Queen Mary, University of London, London, E1 4NS, UK\\ \vspace*{-0.5\baselineskip}
$^{4}$Institute of Astronomy, University of Cambridge, Madingley Road, Cambridge, CB3 0HA, UK
}}}

\date{Accepted XXX. Received YYY; in original form ZZZ}

\pubyear{\the\year{}}

\begin{document}
\label{firstpage}
\pagerange{\pageref{firstpage}--\pageref{lastpage}}
\maketitle

\begin{abstract}
We investigate the impact of Mutual Coupling (MC) between antennas on the time-delay power spectrum response of the core of the SKA-Low radio telescope. Using two in-house tools - FAST (a fast full-wave electromagnetic solver) and OSKAR (a GPU-accelerated radio telescope simulator) - we simulate station beams and compute visibilities for various station layouts (regular, sunflower, and random). Simulations are conducted in an Epoch of Reionisation subband between 120–150~MHz, with a fine spectral resolution of 100~kHz, enabling the investigation of longer delays. Our results show that MC effects significantly increase foreground leakage into longer delays, especially for regular station layouts. For 21-cm science, foreground spill-over into the 21-cm window extends beyond $k_{\parallel} \sim 2$~h$^{-1}$Mpc for all station layouts and across all $k_{\perp}$ modes, completely obscuring the detection window. We find that attempting to remove the foreground contribution from the visibilities using an approximated beam model, based on the average embedded element pattern or interpolating the embedded element patterns from a coarse channel rate of $781$~kHz, results in residuals around $1\%$ ($\sim 10^{11}~\mathrm{mK}^2$h$^{-3}\mathrm{Mpc}^3$)  which is still $7$ orders of magnitude brighter than the expected level of the EoR signal ($\sim 10^{4}~\mathrm{mK}^2$h$^{-3}\mathrm{Mpc}^3$). We also find that station beam models with at least 4-5 significant digits in the far-field pattern and high spectral resolution are needed for effective foreground removal. Our research provides critical insights into the role of MC in SKA-Low experiments and highlights the computational challenges of fully integrating array patterns that account for MC effects into processing pipelines.
\end{abstract}


\section{Introduction}
Over the past decade, the detection of the faint cosmological 21-cm signal, arising from the hyperfine transition of neutral hydrogen atoms, has received increasing attention as one of the frontiers of observational radio astronomy. The signal promises a rich understanding of astrophysics, ranging from the earliest epochs of the Universe (the Dark Ages), through the first luminous sources (the Cosmic Dawn), and the eventual ionization of the intergalactic medium (Epoch of Reionization, or EoR). Measuring the spatial fluctuations of this signal would unlock a deeper understanding of the interplay between different heating and ionizing sources in the early Universe. The primary obstacle preventing such a detection is the isolation of the \mbox{21-cm} signal from astrophysical foregrounds, such as Galactic synchrotron emission, radio galaxies, pulsars, and supernova remnants, which are up to five orders of magnitude brighter than the signal itself \citep{jelic2008foreground}. Foreground avoidance techniques, which exploit the rapid spectral variations of faint signals, enable their isolation at higher delays, helping to separate them from foreground emissions \citep{trott2012impact, parsons2012per, liu2014epocha, liu2014epochb, thyagarajan2015foregrounds, chapman2016effect}. Alternatively, foreground removal techniques \citep{bowman2006sensitivity, liu2009improved, bonaldi2015foreground, chapman2016effect, pober2016importance} aim to subtract modelled foreground contributions from the observed data using knowledge of the telescope's response and the celestial distribution of the foregrounds.
In both foreground avoidance and removal approaches, precise models of the instrument’s beams and analogue chains are essential. Typically, two fundamental questions related to these models must be addressed:
\vspace{-0.1cm}
\begin{enumerate}
    \item How accurately must we model the instrument's response? This focuses on determining the necessary level of accuracy in matching the real instrument response with the model to ensure reliable data interpretation.
    \item In which regions of the observational domain are instrumental effects weakest? Identifying these regions will enable us to focus on areas of minimal instrumental uncertainty to preserve the data quality.
\end{enumerate}
\vspace{-0.1cm}
In addressing the accuracy criterion (question (i)), forecasts of the 21-cm power spectrum from CHIME \citep{bandura2014canadian} indicate that achieving a 0.1\% accuracy in the primary beamwidth is necessary for reliable measurements \citep{shaw2015coaxing, amiri2022overview}. The REACH experiment \citep{de2022reach} has similarly determined that uncertainties in the antenna directivity must remain below $-40$dB to ensure robust global 21-cm detection \citep{cumner2024effects}. The MeerKAT Radio Telescope \citep{jonas2016meerkat} has evaluated the required pointing accuracy to use beam patterns with 1\% accuracy in power \citep{jonas2016meerkat, de2022meerkat}. For SKA-low \citep{5136190}, preliminary studies \citep{nasirudin2022characterizing} indicate that inaccuracies in beam knowledge - resulting from small positional deviations and up to 5\% of broken antennas - can produce residual errors as high as 1\% in the power spectrum EoR window and around 10\% in the foreground wedge. Similarly, for MWA \citep{tingay2013murchison}, 15-40\% of tiles typically have at least one broken dipole, leading to beam modelling errors that produce foreground residuals two orders of magnitude above the 21-cm signal after calibration \citep{10.1093/mnras/stz3375}. With such error levels, systematic artefacts could well be misinterpreted as 21-cm structures i.e. \citet{barry2016, thyagarajan2016effects, ewall2017impact}.

For radio telescopes composed of densely packed antennas, addressing questions (i) and (ii) is complicated by a major issue: Mutual Coupling (MC) \citep{craeye2011review}. In situ, strong electromagnetic interactions between antennas can induce sharp angular and spectral variations in the radiated far field, known as Embedded Element Patterns (EEPs) \citep{bird2021mutual, 10.1117/1.JATIS.8.1.011023, 10501737}. Since each antenna operates in a distinct local environment within the array, the EEPs can vary significantly across elements in the station. In other words, MC effects are direction-dependent, frequency-dependent, and baseline-dependent. To date, MC is identified as a systematic source that potentially limits interferometric 21-cm power spectrum experiments such as LOFAR \citep{van2013lofar, mertens2020improved}, HERA \citep{deboer2017hydrogen, kern2019mitigating, Kern_2020}, PAPER \citep{parsons2010precision}, and MWA \cite{beardsley2016first}. 
LOFAR, NenuFAR \citep{2012sf2a.conf..687Z}, and MWA have implemented full-Jones directional-dependent sky-based calibration algorithms SAGECAL \citep{kazemi2011radio, yatawatta2012gpu}, DDECAL \citep{2018ascl.soft04003V, gan2023assessing} and RTS \citep{4703504}.
Incorporating realistic (simulated or measured) beam models into these calibration pipelines can help to reduce residual errors \citep{Ravi,10.1093/mnras/stae2264}. An alternative approach is mitigating MC effects using fringe rate filters. In this context, the HERA Collaboration modelled first-order dish-dish interactions with a semi-analytical formalism \citep{josaitis2022array, rath2024investigating} to demonstrate the impact of MC at non-zero fringe rates. This informed the development of mitigation strategies and tailored fringe rate filters for foreground avoidance, building on \citet{parsons2016optimized}.

The Square Kilometre Array Low (SKA-Low), examined in this paper, is an interferometric array under construction in Inyarrimanha Ilgari Bundara, Western Australia. It will consist of 512 stations, each containing 256 dual-polarisation SKALA antennas within a 19m radius, spanning $\sim74$km. Mutual coupling (MC) between SKALA antennas inside a station has been a long-standing concern \citep{de2012skala}, with variations in EEPs observed experimentally \citep{6546356, de2018antenna} and through simulations \citep{7297311, 7731490}. Recent UAV measurements \citep{Ravi-NB, 10168937} and electromagnetic simulations \citep{9232307, 10.1117/1.JATIS.8.1.011017, 10501737} confirm that MC introduces notches at frequencies (55MHz, 78MHz, 125~MHz) within the EoR band (50-200 MHz). A recent milestone has also been achieved by incorporating full-wave EEP simulations into calibration algorithms and demonstrating a significant reduction in visibility residuals \citep{Ravi}. Numerical simulations are thus important for understanding the impact of these MC effects on the delay power spectrum of the SKA-Low telescope.

Accurately modelling SKA-low beams requires full-wave simulations of the electromagnetic response of each of the 256 antennas embedded in large (19~meter radius), dense, and irregular arrays across a wide bandwidth (50-350~MHz) with high spectral resolution (below 1~MHz). The complex antenna geometry, small spacing (0.01 wavelengths at the lowest frequency), and large station size (up to 45 wavelengths at the highest frequency) result in a challenging multi-scale Computational Electromagnetics (CEM) problem. General-purpose solvers can take days or weeks per frequency point, even when accelerated with the Multi-Level Fast Multipole Method \citep[MLFMM,][]{engheta1992fast} on large workstations. To overcome this, state-of-the-art CEM methods tailored for large, irregular arrays have been developed \citep{RobMaaskant, bui2018fast, 10108930, chose2023physics}. The in-house solver, named Fast Array Simulation Tool (FAST) \citep{gueuning2024FAST}, simulates all the antenna responses in the station in about 10 minutes per frequency point on a current standard laptop, with re-simulation for a different station layout taking just 1 minute, enabling fine spectral sampling and therefore evaluation of impulse responses up to late time delays.


Modelling SKA-low visibilities also presents computational challenges. Each of the 512 stations generates beamformed signals, which are then cross-correlated to produce over $130,000$ visibility pairs. To ensure accuracy on SKA-low's longest baselines (up to $\sim 74$~km), high-resolution sky maps with tens of millions of pixels are essential for avoiding aliasing in the sky integrals. Furthermore, each station's beam, derived from precomputed EEP data, must be re-evaluated across all these sky directions. To address this, the tool OSKAR has been developed \cite{dulwich_fred_2020_3758491} to leverage GPU parallelization, and we have also implemented a GPU-parallelized approach for station beam evaluation, originally proposed in \citet{bui2018fast}.

In this paper, we use the in-house tools, FAST and OSKAR, to investigate the accuracy required for the EEP models (question (i)) as well as the impact of MC effects on the time-delay power spectrum of foreground emissions (question (ii)) through numerical simulations. To address question (i), we estimate delay power spectrum residuals after foreground removal, using an approximate station beam model that is corrupted by random noise, or based on the Average embedded Element Pattern (AEP) \citep{wijnholds2019using} or spectrally-interpolated EEPs. Regarding question (ii), we investigate MC effects by comparing beam and visibility time-delay responses across three station layouts - regular, sunflower, and random. We quantify foreground leakage into longer delays caused by MC, considering both rotated and non-rotated stations. This paper also highlights the computational costs associated with fully integrating array patterns, which fully account for MC effects, in the SKA-Low pipeline. This paper is organized as follows. Section~\ref{sec:conv} defines the conventions for the coordinates systems used, while Section~\ref{sec:formulism} presents the mathematical formulation for the forward model of these visibilities, based on EEPs, array patterns, and sky brightness distribution. In Section~\ref{sec:simparameters}, we outline the parameters used for the telescope simulations. Section~\ref{sec:pipeline} describes our simulation pipeline, emphasizing the recent advancements introduced in the tools, FAST and OSKAR. Section~\ref{sec:numresults} presents the simulation results, including an analysis of the station beam impulse response across different layouts, the foreground spill-over induced by MC, and the residual levels after foreground removal using several approximate beam models. Finally, we summarize our findings in Section~\ref{sec:conclusion}.

\section{Conventions}
\label{sec:conv}
Before outlining our mathematical derivations, we would like to clarify the notation used by the coordinate systems in this paper.

Comoving distances are spatial coordinates that measure separation in cosmology. Per their definition, these distances remain constant over time despite the expansion of the universe, as they account for the changing scale at different epochs. Assuming a Cartesian coordinate system with unit vectors ($\hat{\mathbfit{x}},\hat{\mathbfit{y}},\hat{\mathbfit{z}}$), the comoving coordinates will be denoted as follows:
\begin{equation}
\mathbfit{r} = r_x \hat{\mathbfit{x}}+ r_y \hat{\mathbfit{y}}+ r_z \hat{\mathbfit{z}}.
\end{equation}
The spatial frequency coordinates, which are associated with the cosmological Fourier domain and referred to as the $k$-space, are given by:
\begin{equation} 
\label{kvec}
\mathbfit{k} = k_x \hat{\mathbfit{x}} + k_y \hat{\mathbfit{y}} + k_z \hat{\mathbfit{z}},
\end{equation} 
where $\mathbfit{k}$ is the wavevector with cosmological wavenumber $k=||\mathbfit{k}||$.

Physical relative distances, known as baselines, are represented by the baseline vector $\mathbfit{b}$:
\begin{equation}
\mathbfit{b} = u \hat{\mathbfit{x}}+ v \hat{\mathbfit{y}}+ w \hat{\mathbfit{z}},
\end{equation}
when expressed in meters, or by $\mathbfit{b}_\lambda = \mathbfit{b} /\lambda_0$ when expressed in units of the free-space wavelength $\lambda_0$. The vector $\mathbfit{u}_\lambda =  u/\lambda_0 \,\hat{\mathbfit{x}}+ v/\lambda_0\, \hat{\mathbfit{y}}$ represents the in-plane projection of the baseline vector $\mathbfit{b}_\lambda$. In this paper, the baseline notation $\mathbfit{b}$ refers to the separation between station centres, while $\mathbfit{b}_{ant}$ (in meters) denotes the position of an antenna with respect to its station centre. 

The directional unit vector $\hat{\mathbfit{s}}$, which represents a line of sight, is given by:
\begin{equation}
\hat{\mathbfit{s}} = l \hat{\mathbfit{x}}+ m \hat{\mathbfit{y}}+ n \hat{\mathbfit{z}},
\end{equation}
where $n = (1-l^2-m^2)^{1/2}$, ensuring that $\hat{\mathbfit{s}}$ is constrained to unit-length.

\section{A Forward model of telescope visibilities}
\label{sec:formulism}
A \textit{forward model} simulates how signals from a celestial distribution are transformed by the instrument, capturing the distortions introduced at each stage of the system. This model is essential for understanding how instrumental effects propagate through the digital processing pipeline and manifest in the final product, the delay power spectrum. In this section, we focus on modelling the measured visibilities, specifically including the impact of MC between antennas in the beam response of each station.

\subsection{System overview} 
\begin{figure}
\centering
\begin{tikzpicture}[scale=\textwidth/22cm,samples=200]
\draw [black,very thick] (0-1.55-1.1,0.0) -- (0-1.55-1.1,0.5);
\draw [black,very thick] (0-1.55-1.1,0.5) -- (0.5-1.55-1.1,1.0);
\draw [black,very thick] (0-1.55-1.1,0.5) -- (-0.5-1.55-1.1,1.0);
\draw [black,very thick] (0-1.55-2*1.1,0.0) -- (0-1.55-2*1.1,0.5);
\draw [black,very thick] (0-1.55-2*1.1,0.5) -- (0.5-1.55-2*1.1,1.0);
\draw [black,very thick] (0-1.55-2*1.1,0.5) -- (-0.5-1.55-2*1.1,1.0);
\draw [black,very thick] (0-1.55-3*1.1,0.0) -- (0-1.55-3*1.1,0.5);
\draw [black,very thick] (0-1.55-3*1.1,0.5) -- (0.5-1.55-3*1.1,1.0);
\draw [black,very thick] (0-1.55-3*1.1,0.5) -- (-0.5-1.55-3*1.1,1.0);
\draw [black] (-1.55-1.1,0.0) -- (-1.55-1.1,-2.0+0.75) ;
\draw [black] (-1.55-2*1.1,0.0) -- (-1.55-2*1.1,-2.0+0.75) ;
\draw [black] (-1.55-3*1.1,0.0) -- (-1.55-3*1.1,-2.0+0.75) ;
\draw [black] (-1.55-1.1,-2.0+0.75) -- (-1.55-2*1.1,-2.0);
\draw [black] (-1.55-2*1.1,-2.0+0.75) -- (-1.55-2*1.1,-2.0);
\draw [black] (-1.55-3*1.1,-2.0+0.75) -- (-1.55-2*1.1,-2.0);

\draw [draw=black] (-1.95,-2.5) rectangle (-5.6,-0.4);
\draw [ black,fill=white] (-1.55-3*1.1,-2.0+0.75)  circle (0.15cm);
\draw [black] (-1.55-3*1.1-0.1061,-2.0+0.1061+0.75) -- (-1.55-3*1.1+0.1061,-2.0-0.1061+0.75);
\draw [black] (-1.55-3*1.1+0.1061,-2.0+0.1061+0.75) -- (-1.55-3*1.1-0.1061,-2.0-0.1061+0.75);
\draw [ black,fill=white] (-1.55-2*1.1,-2.0+0.75)  circle (0.15cm);
\draw [black] (-1.55-2*1.1-0.1061,-2.0+0.1061+0.75) -- (-1.55-2*1.1+0.1061,-2.0-0.1061+0.75);
\draw [black] (-1.55-2*1.1+0.1061,-2.0+0.1061+0.75) -- (-1.55-2*1.1-0.1061,-2.0-0.1061+0.75);
\draw [ black,fill=white] (-1.55-1.1,-2.0+0.75)  circle (0.15cm);
\draw [black] (-1.55-1.1-0.1061,-2.0+0.1061+0.75) -- (-1.55-1.1+0.1061,-2.0-0.1061+0.75);
\draw [black] (-1.55-1.1+0.1061,-2.0+0.1061+0.75) -- (-1.55-1.1-0.1061,-2.0-0.1061+0.75);

\node[text width=0.5cm,align=left] at (-1.55-3*1.1-0.39,-0.75)
{$w_{11}$};
\node[text width=0.5cm,align=left] at (-1.55-2*1.1-0.39,-0.75)
{$w_{12}$};
\node[text width=0.5cm,align=left] at (-1.55-1.1-0.39,-0.75)
{$w_{13}$};

\node[text width=0.5cm, rotate=0, align=left] at (-1.0-3*1.1-0.45+0.5,1.5)
{$v_{\textrm{ant},n1}(f,t)$};

\draw [black,->] (-1.55-2*1.1,-2.0) -- (0.5-1.675,-4.5);
\draw [ black,fill=white] (-1.55-2*1.1,-2.0) circle (0.15cm);
\draw [black] (-1.55-2*1.1-0.15,-2.0) -- (-1.55-2*1.1+0.15,-2.0);
\draw [black] (-1.55-2*1.1,-2.0-0.15) -- (-1.55-2*1.1,-2.0+0.15);

\node[text width=0.5cm,align=left] at (-5.25,-2.25)
{$v_1(f, t)$};

\draw [black,very thick] (5.0+0-1.55-1.1,0.0) -- (5.0+0-1.55-1.1,0.5);
\draw [black,very thick] (5.0+0-1.55-1.1,0.5) -- (5.0+0.5-1.55-1.1,1.0);
\draw [black,very thick] (5.0+0-1.55-1.1,0.5) -- (5.0+-0.5-1.55-1.1,1.0);
\draw [black,very thick] (5.0+0-1.55-2*1.1,0.0) -- (5.0+0-1.55-2*1.1,0.5);
\draw [black,very thick] (5.0+0-1.55-2*1.1,0.5) -- (5.0+0.5-1.55-2*1.1,1.0);
\draw [black,very thick] (5.0+0-1.55-2*1.1,0.5) -- (5.0+-0.5-1.55-2*1.1,1.0);
\draw [black,very thick] (5.0+0-1.55-3*1.1,0.0) -- (5.0+0-1.55-3*1.1,0.5);
\draw [black,very thick] (5.0+0-1.55-3*1.1,0.5) -- (5.0+0.5-1.55-3*1.1,1.0);
\draw [black,very thick] (5.0+0-1.55-3*1.1,0.5) -- (5.0+-0.5-1.55-3*1.1,1.0);
\draw [black] (5.0+-1.55-1.1,0.0) -- (5.0+-1.55-1.1,-2.0+0.75) ;
\draw [black] (5.0+-1.55-2*1.1,0.0) -- (5.0+-1.55-2*1.1,-2.0+0.75) ;
\draw [black] (5.0+-1.55-3*1.1,0.0) -- (5.0+-1.55-3*1.1,-2.0+0.75) ;
\draw [black] (5.0+-1.55-1.1,-2.0+0.75) -- (5.0+-1.55-2*1.1,-2.0);
\draw [black] (5.0+-1.55-2*1.1,-2.0+0.75) -- (5.0+-1.55-2*1.1,-2.0);
\draw [black] (5.0+-1.55-3*1.1,-2.0+0.75) -- (5.0+-1.55-2*1.1,-2.0);

\draw [draw=black] (5.0+-1.95,-2.5) rectangle (5.0+-5.8,-0.4);
\draw [ black,fill=white] (5.0+-1.55-3*1.1,-2.0+0.75)  circle (0.15cm);
\draw [black] (5.0+-1.55-3*1.1-0.1061,-2.0+0.1061+0.75) -- (5.0+-1.55-3*1.1+0.1061,-2.0-0.1061+0.75);
\draw [black] (5.0+-1.55-3*1.1+0.1061,-2.0+0.1061+0.75) -- (5.0+-1.55-3*1.1-0.1061,-2.0-0.1061+0.75);
\draw [ black,fill=white] (5.0+-1.55-2*1.1,-2.0+0.75)  circle (0.15cm);
\draw [black] (5.0+-1.55-2*1.1-0.1061,-2.0+0.1061+0.75) -- (5.0+-1.55-2*1.1+0.1061,-2.0-0.1061+0.75);
\draw [black] (5.0+-1.55-2*1.1+0.1061,-2.0+0.1061+0.75) -- (5.0+-1.55-2*1.1-0.1061,-2.0-0.1061+0.75);
\draw [ black,fill=white] (5.0+-1.55-1.1,-2.0+0.75)  circle (0.15cm);
\draw [black] (5.0+-1.55-1.1-0.1061,-2.0+0.1061+0.75) -- (5.0+-1.55-1.1+0.1061,-2.0-0.1061+0.75);
\draw [black] (5.0+-1.55-1.1+0.1061,-2.0+0.1061+0.75) -- (5.0+-1.55-1.1-0.1061,-2.0-0.1061+0.75);

\node[text width=0.5cm,align=left] at (5.0+-1.55-3*1.1-0.39,-0.75)
{$w_{21}$};
\node[text width=0.5cm,align=left] at (5.0+-1.55-2*1.1-0.39,-0.75)
{$w_{22}$};
\node[text width=0.5cm,align=left] at (5.0+-1.55-1.1-0.39,-0.75)
{$w_{23}$};

\node[text width=0.5cm, rotate=0, align=left] at (5+-1.0-3*1.1-0.45+0.5,1.5){$v_{\textrm{ant},n2}(f,t)$};

\draw [black,->] (5.0+-1.55-2*1.1,-2.0) -- (0.5-1.675,-4.5);
\draw [ black,fill=white] (5.0+-1.55-2*1.1,-2.0) circle (0.15cm);
\draw [black] (5.0+-1.55-2*1.1-0.15,-2.0) -- (5.0+-1.55-2*1.1+0.15,-2.0);
\draw [black] (5.0+-1.55-2*1.1,-2.0-0.15) -- (5.0+-1.55-2*1.1,-2.0+0.15);

\node[text width=0.5cm,align=left] at (5.3+-3.25,-2.25)
{$v_2(f, t)$};

\draw [black,->] (0.5-1.675,-4.5) -- (0.5-1.675,-6.7);
\node[text width=0.5cm,align=left] at (0.5-1.8,-7.2)
{$V_{12}(f)$};

\draw [draw=black] (2.5+-2.0,-6.2) rectangle (2.5+-5.35,-3.5);

\draw [ black,fill=white] (0.5-1.675,-4.5)  circle (0.15cm);
\draw [black] (-0.1061-1.175,-4.5+0.1061) -- (0.1061-1.175,-4.5-0.1061);
\draw [black] (0.1061-1.175,-4.5+0.1061) -- (-0.1061-1.175,-4.5-0.1061);

\draw [draw=black,fill=white] (0.9-1.675,-3.8) rectangle (0.9-1.675+0.3,-4.1);
\node[text width=0.25cm,align=left] at (0.9-1.675+0.2,-3.8-0.15) 
{$*$};
\draw [draw=black,fill=white] (-0.35-1.675+0.5,-5.5-0.35) rectangle (0.35-1.675+0.5,-5.5+0.35);
\node[text width=0.25cm,align=left] at (-1.675+0.5,-5.5)
{$\int$};

\draw [black,dotted] (-7.0,0.0) -- (5.0-2.1,0.0);

\draw [black,dotted] (-7.0,-3.0) -- (5.0-2.1,-3.0);

\node[text width=2.0cm,align=left,rotate=90] at (-6.5,1.5)
{\footnotesize RF SYSTEM};

\node[text width=2.5cm,align=left,rotate=90] at (-6.5,-1.3)
{\footnotesize BEAMFORMERS};

\node[text width=2.5cm,align=left,rotate=90] at (-6.5,-4.5)
{\footnotesize CORRELATORS};

\node[text width=2.0cm,align=left] at (-3.5,2.9)
{\footnotesize STATION 1};

\node[text width=2.0cm,align=left] at (1.5,2.9)
{\footnotesize STATION 2};

\end{tikzpicture}
\caption{Block diagram of a radio telescope composed of phased array stations. Channelized voltages $v_{\textrm{ant},np}(f, t)$ from each antenna are beamformed within each station $p$, and the resulting beamformed voltages are then cross-correlated.}
\label{fig:beamformed}
\end{figure}
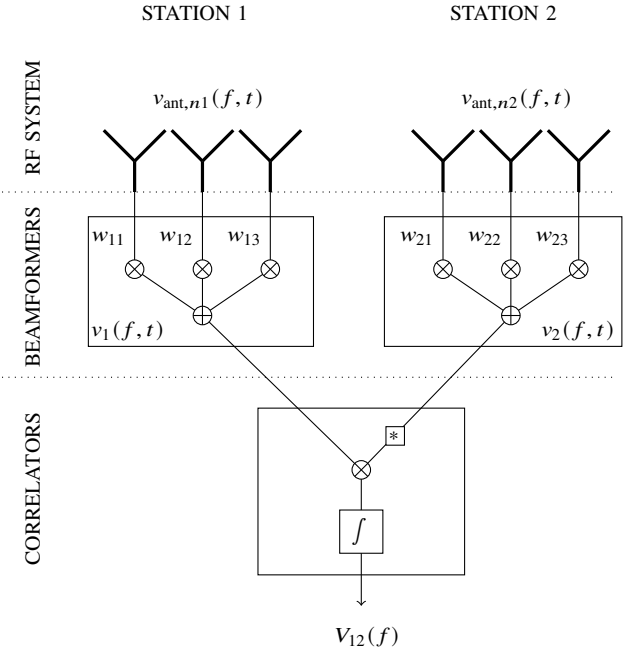

The interferometer setup comprises $N_s$ phased array stations, each containing $N_a$ antennas. After digitization and channelization of raw voltages from each antenna, the voltage from antenna $n$ in station $p$ is represented as a time-varying mono-frequency signal $v_{\textrm{ant},np}(t, f)$. The $N_a$ voltages in each station are then beamformed as $v_{p}(f,t) = \sum_n w_{np}(f) v_{\textrm{ant},np}(f,t)$, with beamforming weights $w_{np}(f)$ chosen to steer the beam or control its shape. Beamformed voltages are then cross-correlated to yield the complex visibility $V_{pq}(f) = V(\mathbfit{b}_{\lambda,pq}, f)$, where $\mathbfit{b}_{\lambda,pq}$ is the baseline vector linking the centres of stations $p$ and $q$. A block diagram for a two-station, three-antenna configuration is shown in Fig.~\ref{fig:beamformed}.

\subsection{Mathematical model of the visibilities} 
The following derivations outline how the visibilities, measured by cross-correlating beamformed signals from different stations, are modelled using both celestial distribution and beam models.

Let us first assume a monochromatic incident plane wave propagating in the direction $\hat{\mathbfit{s}}$, with a time-varying electric field vector 
$\mathbfit{E}_{i}(f,t)$ representing its amplitude and polarisation, where the subscript $i$ denotes incident. Assuming and suppressing the time dependence $e^{j2\pi ft}$, the electrical field of the incident plane wave is expressed as:
\begin{equation} \label{incident_field}
\mathbfit{E}(f,t,\mathbfit{b}) = \mathbfit{E}_i(f,t) \ e^{j k_0\hat{\mathbfit{s}} \cdot \mathbfit{b}},
\end{equation}
where $k_0$ is the free-space wavenumber. First, we consider single-polarised antennas (with a single feed port). According to Lorentz’s reciprocity theorem, the voltage $v_{\textrm{ant},np}(f,t)$ appearing at the output port of a passive antenna $n$ belonging to a station $p$ illuminated by the plane wave can be obtained by treating antenna $n$ as active while all the other antennas are passively terminated \citep{kildal2015foundations}. Using the centre of antenna $n$ as a phase reference in \eqref{incident_field}, this voltage is given by \citep{craeye2011review}:
\begin{equation} \label{antenna_voltage}
    v_{\textrm{ant},np}(f,t) = \dfrac{2\lambda_0}{j \eta_0} \ Z_{L,np}(f) \ \mathbf{ EEP}_{np}(f,\hat{\mathbfit{s}}) \cdot \mathbfit{E}_{i} (f,t),
\end{equation}
where $\eta_0$ is the free-space characteristic impedance, $Z_{L,np}$ is the input impedance of the amplifier connected to antenna $n$. The EEP, $\mathbf{EEP}_{np}$ in  
\eqref{antenna_voltage}, is expressed in volts and evaluated in the active case by feeding antenna $n$ with a Thevenin equivalent of impedance $Z_{L,np}$ and a source of unit voltage. In the passive case described in \eqref{antenna_voltage}, the EEP is implicitly normalized by 1 volt, so that it becomes a dimensionless quantity. From there, the beamformed voltage associated with station $p$ can be expressed as:
\begin{equation}
\label{beam}
v_{p}(f, t) = \dfrac{2\lambda_0}{j \eta_0} \ Z_0 \  \mathbf{AP}_{p}(f,\hat{\mathbfit{s}}) \cdot \mathbfit{E}_i (f,t),
\end{equation}
where the Array Pattern (AP) is defined by:
\begin{equation}
\label{AP}
\mathbf{AP}_p(f,\hat{\mathbfit{s}}) = \sum_{n=1}^{N_a} c_{pn}(f) \  \mathbf{EEP}_{np} (f,\hat{\mathbfit{s}}) \ e^{j k_0  \hat{\mathbfit{s}} \cdot \mathbfit{b}_{\textrm{ant},np}},
\end{equation}
with the coefficients $c_{pn} = w_{pn} Z_{L,np}/Z_0 $. The phase reference is now taken at the centre of the station, introducing a phase factor in \eqref{AP} based on the relative antenna distances $\mathbfit{b}_{\textrm{ant},np}$. The impedance $Z_0$ in \eqref{beam} has been introduced so that the AP shares the same units as the EEPs. 

Considering now dual-polarised antennas with two feed ports, denoted as $v_{p,X}$ and $v_{p,Y}$, with indices $X$ and $Y$ referring to the orientation of the associated dipoles along $\hat{\mathbfit{x}}$ or $\hat{\mathbfit{y}}$ axis, we re-express the beamformed voltages using the $2\times2$ Jones matrix formalism \cite{Jones:41, smirnov2011revisiting}. Omitting the explicit time ($t$) and frequency ($f$) dependence, this yields:
\begin{equation}
 \begin{bmatrix} v_{p,X} \\ v_{p,Y} \end{bmatrix} =  \begin{bmatrix} J_{p,XX} & J_{p,XY} \\ J_{p,YX} & J_{p,YY} \end{bmatrix}  \begin{bmatrix} E_{i,X} \\ E_{i,Y}\end{bmatrix}
\end{equation}
where $\mathbfit{E}_i = E_{i,X}\,\hat{\mathbfit{e}}_{X} + E_{i,Y}\,\hat{\mathbfit{e}}_{Y}$, with $\hat{\mathbfit{e}}_X$ and $\hat{\mathbfit{e}}_Y$ representing the unit vectors corresponding to the horizontal and vertical components in the Ludwig-III polarisation coordinate system \cite{1140406}. In agreement with our previous derivation \eqref{beam}, the elements of the Jones matrix are given by:
\begin{equation}
\begin{bmatrix} J_{p,XX} & J_{p,XY} \\ J_{p,YX} & J_{p,YY} \end{bmatrix} = \dfrac{2\lambda_0 }{j \eta_0} \ Z_0 \
  \begin{bmatrix}
   \textrm{AP}_{p,XX} &
   \textrm{AP}_{p,XY} \\
   \textrm{AP}_{p,YX} &
   \textrm{AP}_{p,YY}
   \end{bmatrix}
\end{equation}
where $\mathbf{AP}_{p,X} = \textrm{AP}_{p,XX}\,\hat{\mathbfit{e}}_{X} + \textrm{AP}_{p,XY}\,\hat{\mathbfit{e}}_{Y}$ and $\mathbf{AP}_{p,Y} = \textrm{AP}_{p,YX}\,\hat{\mathbfit{e}}_{X} + \textrm{AP}_{p,YY}\,\hat{\mathbfit{e}}_{Y}$ are the APs for feed ports $X$ and $Y$, respectively. This formalism allows us to apply the Radio Interferometric Measurement Equation (RIME), as defined in \cite{hamaker1996understanding1, smirnov2011revisiting}, to relate the Stokes I parameter of the measured visibilities $V_{pq}$ to the spectral brightness distribution $B(f,\hat{\mathbfit{s}})$ and to the station beams:
\begin{align}
\label{Zernike}  
    V_{pq} (f,\mathbfit{b}_\lambda) = \iint A_{pq}(f,\hat{\mathbfit{s}}) \ B(f,\hat{\mathbfit{s}})\ e^{-j 2\pi \hat{\mathbfit{s}} \cdot \mathbfit{b}_{\lambda,pq}} \  \frac{\mathrm{d}l}{n} \mathrm{d}m,
\end{align}
with the instrument response: 
\begin{align} \nonumber
\label{AP_product}
A_{pq}(f,\hat{\mathbfit{s}}) & =  \frac{1}{\Omega_{pq} (f)} \ \bigg(J_{p,XX} \ J_{q,XX}^* \ +  \ J_{p,XY} \ J_{q,XY}^* \\ &  + \ J_{p,YX} \ J_{q,YX}^* \ + \ J_{p,YY} \ J_{q,YY}^* \bigg) ,
\end{align}
where $*$ denotes the complex conjugate. We refer to this function $A_{pq}(f,\hat{\mathbfit{s}})$ as the \textit{Beam Transfer Function} (BTF) while its time-delay Fourier counterpart $\tilde{A}_{pq}(\tau,\hat{\mathbfit{s}})$ is called the \textit{Beam Impulse Response} (BIR). The factor $\Omega_{pq}$ is called the integrated beam response and is given by:
\begin{equation}
\label{Omegapq}
\Omega_{pq}(f) = \iint |A_{pq}(f,\hat{\mathbfit{s}})| \ \frac{\mathrm{d}l}{n} \ \mathrm{d}m
\end{equation}
This factor ensures that the visibilities $V_{pq}$ are expressed in Jy units when the brightness $B$ is expressed in Jy/sr. To alleviate the notations, we will temporarily assume beams similar across stations, making the BTF effectively baseline-independent, that is $A_{pq} = A$.

\subsection{Time-delay transform} 
\label{subsec:timedelaytransform}
As explained in Appendix~\ref{sec:DPS}, the power spectrum $\Tilde{\mathcal{P}}$ of an astronomical signal can be estimated from the amplitude of the time-delay Fourier transform $\tilde{V}(\tau,\mathbfit{b})$, expressed in JyHz, of the visibilities $V(f,\mathbfit{b}_\lambda)$. In practice, the visibilities are only measured within a limited sub-band $(f_{\min},f_{\max})$, a subset of the telescope’s full operational range. To minimize truncation effects, a window function $W(f)$ is applied to the visibilities, resulting in:
\begin{equation}
\label{est_delaytransform}
    \Tilde{V}(\tau,\mathbfit{b}) = \int_{f_{\min}}^{f_{\max}}  W(f) \ V(f,\mathbfit{b}_\lambda) \ e^{-2 \pi j f \tau} \ \mathrm{d}f.
\end{equation} 
Substituting \eqref{Zernike} into \eqref{est_delaytransform} and applying the convolution theorem, this expression becomes:
\begin{equation}
\label{delay_transform}
    \Tilde{V}(\tau,\mathbfit{b}) = \tilde{W}(\tau) \ast \iint \tilde{A}(\tau,\hat{\mathbfit{s}}) \ast \tilde{B}(\tau - \hat{\mathbfit{s}} \cdot \mathbfit{b}/c_0,\hat{\mathbfit{s}}) \ \frac{\mathrm{d}l}{n} \mathrm{d}m,
\end{equation}
where $\ast$ denotes a convolution product, $c_0$ is the free-space speed of light, $\tilde{W}(\tau)$ is the time-delay response of the window function $W$ and and the spectrum $\tilde{B}$ is the Fourier transform of the brigthness $B$. The spectral sampling step $\Delta f$ is a critical factor in computing the delay Fourier transform \eqref{est_delaytransform}. For time-limited functions over duration $T$, the Nyquist criterion requires $\Delta f = \pi / (2T)$  to confine aliasing errors outside the observed delay range $\tau \in [0, T]$.

\subsection{Instrumental effects in the baseline-delay domain}
\begin{figure}
\centering
\includegraphics[width=0.9\linewidth]{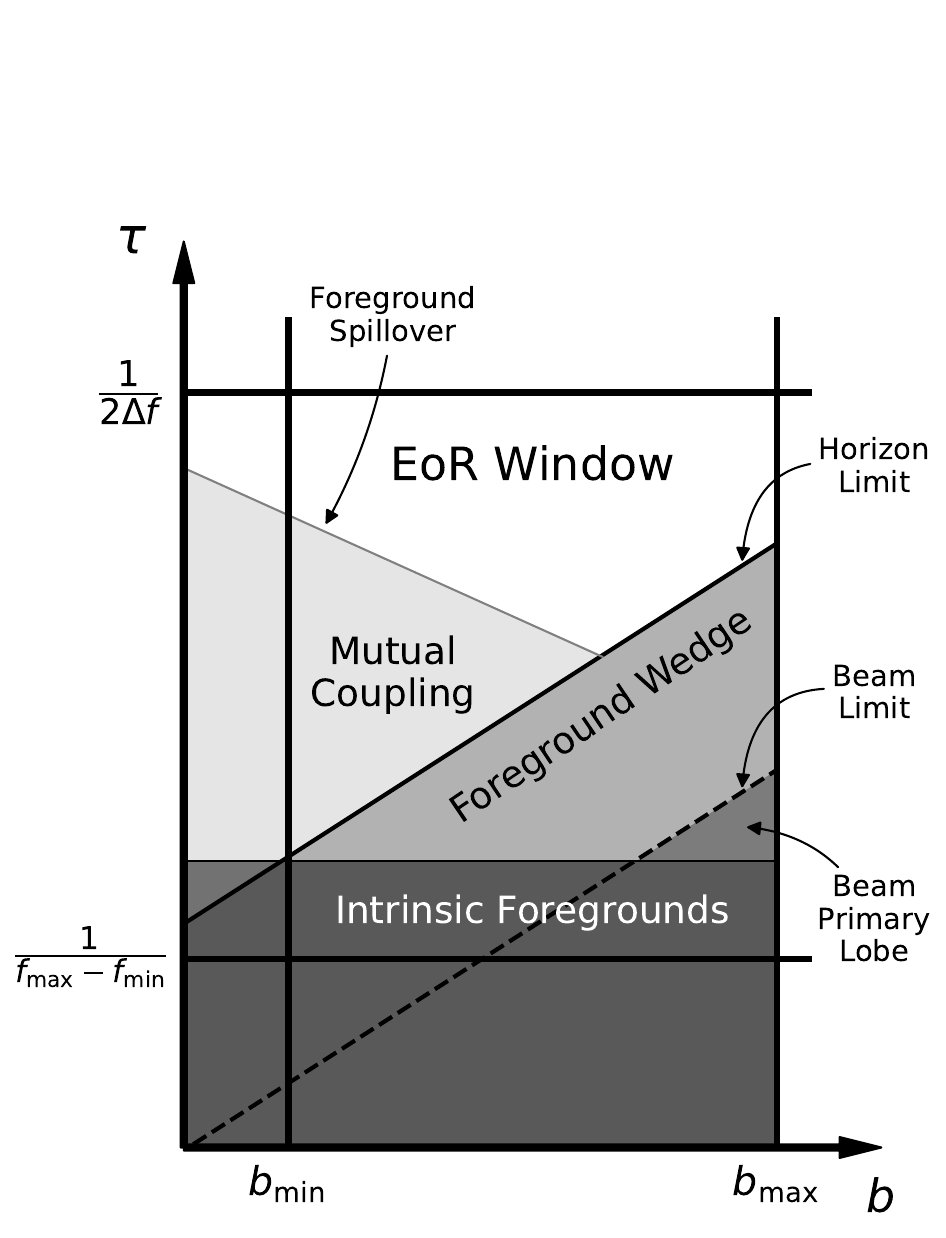}
\caption{A log space illustration of instrumental effects in the delay-baseline domain.}
\label{fig:timedelayeffects}
\end{figure}

Time-delay visibilities are analysed in baseline length-delay space, $(b,\tau)$, by grouping visibilities into baseline length bins and averaging within each bin to obtain the average visibility, $\tilde{V}_{av}(b,\tau)$. 
For simplicity, we consider the case of a single point source denoted by index $i$ of flux density $S(f) = B(f,\hat{\mathbfit{s}}) \Delta l_i \Delta m_i/n_i$ in direction $\hat{\mathbfit{s}}_i$ and solid angle $\Delta l_i \Delta m_i/n_i$. The relation \eqref{delay_transform} thus reduces to:
\begin{equation}
\label{TD_vis_1source}
    \Tilde{V}(\tau,\mathbfit{b}) = \tilde{W}(\tau) \ast \tilde{A}(\tau,\hat{\mathbfit{s}}_i) \ast \tilde{S}(\tau - \hat{\mathbfit{s}}_i \cdot \mathbfit{b}/c_0),
\end{equation}
The effect of the instrument appears as a cascade of convolution products in the time-delay domain, which we will analyse sequentially. All these effects are summarised and sketched in Fig.~\ref{fig:timedelayeffects}.

Firstly, the baseline exponential in \eqref{Zernike} causes a translation of sources flux density spectrum, denoted as $\tilde{S}$, by a time delay that scales with the projection \citep{parsons2012per, chapman2016effect},
\begin{equation}
\label{geo_delay}
\tau = \hat{\mathbfit{s}} \cdot \mathbfit{b}/c_0.
\end{equation}
This relation indicates that, for planar baselines with $\mathbfit{b}=\mathbfit{u}$, sources positioned at the zenith introduce no delay, while sources oriented in line with or opposite to the baseline direction yield either positive or negative delays.
The average visibility $\Bar{V}_{av}(b,\tau)$ peaks around the delay $\tau = b \sin \theta_i /c_0$ with the source zenith angle $\theta_i$, thus following a line with slope $\sin \theta_i/c_0$ in baseline length-delay space ($b$, $\tau$). This delay reaches its maximum, the \textit{horizon limit}, at $\tau=b/c_0$, when the source is on the horizon \citep{morales2012four}. 
As the baseline length $b$ increases, the separation between spectrum peaks at different elevation angles $\theta_i$ increases, enhancing resolution in the time-delay domain. For sources with spatially smooth brightness distribution $B$ (diffuse emission), the visibility $\Bar{V}_{av}(b,\tau)$ decreases rapidly with baseline length $b$ and increases with delay $\tau$ peaking at the horizon limit.

Secondly, the shifted source spectrum $\tilde{S}$ is convolved with the BIR, $\tilde{A}(\tau, \hat{\mathbfit{s}}_i)$, in the direction $\hat{\mathbfit{s}}_i$. Sources within the main lobe thus appear brighter than those in the sidelobes, and higher sidelobe levels manifest in the $(b, \tau)$ plots as increased peak values around $\tau = \sin \theta_i b / c_0$. The beam limit, defined by $\tau = b \sin \theta_i / c_0$, is based on the half-beamwidth $\theta_i$ at the centre frequency. The BIR can decay slowly with delay $\tau$, and when convolved with a more rapidly decaying signal spectrum $\tilde{S}$, such as foreground emission, it spreads the signal power into longer delays, causing \textit{leakage} that can extend far beyond the horizon limit, as shown in Section~\ref{sec:numresults}. This slow decay arises from unwanted electromagnetic interactions, such as multiple reflections or scattering, which prolong wave attenuation. Optimizing antenna geometries and array configurations during design can help minimize the BIR decay rate.

Finally, $\tilde{W}(\tau)$ in \eqref{TD_vis_1source} is the Fourier transform of the frequency window function $W(f)$, used to reduce ringing artefacts from truncating the frequency band to $(f_{\min}, f_{\max})$. A window with a narrow first lobe and high dynamic range is preferred \citep{lanman2020quantifying}; here, we use the convolved Blackman-Harris window as in \citet{parsons2014new}. The first lobe of the combined response $\tilde{W} \ast \tilde{A}$ matches or exceeds that of $\tilde{W}$, depending on beam chromaticity. However, the dynamic range is generally governed by the slower decay of the beam response $\tilde{A}$.

\subsection{Delay Power Spectrum}
After computing the time-delay Fourier transform to the visibilities $V_{pq}$ to time-delay visibilities $\tilde{V}_{pq}$, we can then evaluate the delay power spectrum from these time-delay visibilities using the following formula \citep{morales2004toward}:
\begin{equation}
\Tilde{\mathcal{P}}(k_\perp, k_\parallel) \simeq \dfrac{X^2(z)Y(z)}{\Omega_{A}} \ |\Tilde{V}(\tau,\mathbfit{b})|^{2},
\label{dps}
\end{equation}
where $z$ is the redshift, $X(z)$ and $Y(z)$ are cosmological scaling factors, and $\Omega_A$ is a normalisation factor that depends on the instrument response $A$.  The delay power spectrum is often represented as a function of the parallel and perpendicular components, $\mathbfit{k}_\perp$ and $\mathbfit{k}_\parallel$, of the cosmological wavevector $\mathbfit{k}$ defined in \eqref{kvec}. The link to baseline time-delay coordinates is given by :
\begin{equation}
\mathbfit{k}_\perp = \dfrac{2\pi \mathbfit{u}_\lambda}{X(z)},  \hspace{1cm} k_{\parallel} = \dfrac{2\pi \tau}{Y(z)}.
\label{kparkperp}
\end{equation}
For interested readers, a detailed reminder about the derivation of formulas \eqref{dps} and \eqref{kparkperp}, along with an expansion of the various constants, is provided in Appendix~\ref{sec:DPS}.


\section{Numerical Experiment: SKA-Low}
\label{sec:simparameters}
The spiral configuration of SKA-Low features expanding arms with baseline lengths up to $\sim 74$~km for higher resolution, alongside a core of 224 densely packed stations that provide instantaneous uv-coverage with minimum baseline lengths starting from around $b_{\min} = 40$~m to around $b_{\max} = 1000$~m. One of SKA-Low’s key scientific goals is the detection of the \mbox{21-cm} signal. In this section, we outline the parameters of a realistic scenario considered for the detection of this signal, with results to be presented in Section~\ref{sec:numresults}.

\subsection{Observation parameters} 
The SKA Epoch of Reionisation Science Working Group has declared their intent to target the 21-cm signal between the redshift of $6<$ z $<30$ \citep{bourke2015advancing} and therefore an approximate frequency range of $46~$MHz $<$ f $<$ $202~$MHz. In this work, we focus on a sub-band from $f_{\text{min}}=120$~MHz to $f_{\text{max}}=150$~MHz at a 100~kHz spectral resolution ($\Delta f$) placing us firmly within the 21-cm band while allowing us to highlight the impact of MC-induced features \citep{9232307, 10501737} present in the beam.
To minimize the ratio between the 21-cm signal and astrophysical foregrounds, observation times are typically chosen to ensure that the beam's primary lobe and the field of view are relatively free from bright foreground emissions. This involves avoiding areas such as the Galactic Centre (GC) and prominent point sources like Rigel. An example of a suitable observation field is the MWA EOR1, located at $\mathrm{RA} = 60^{\circ}$ and $\mathrm{Dec} = -27^{\circ}$ (see \citealt{bowman2013science, beardsley2016first}). 
\begin{figure}
    \begin{subfigure}{\linewidth}
    \includegraphics[width=1\linewidth]{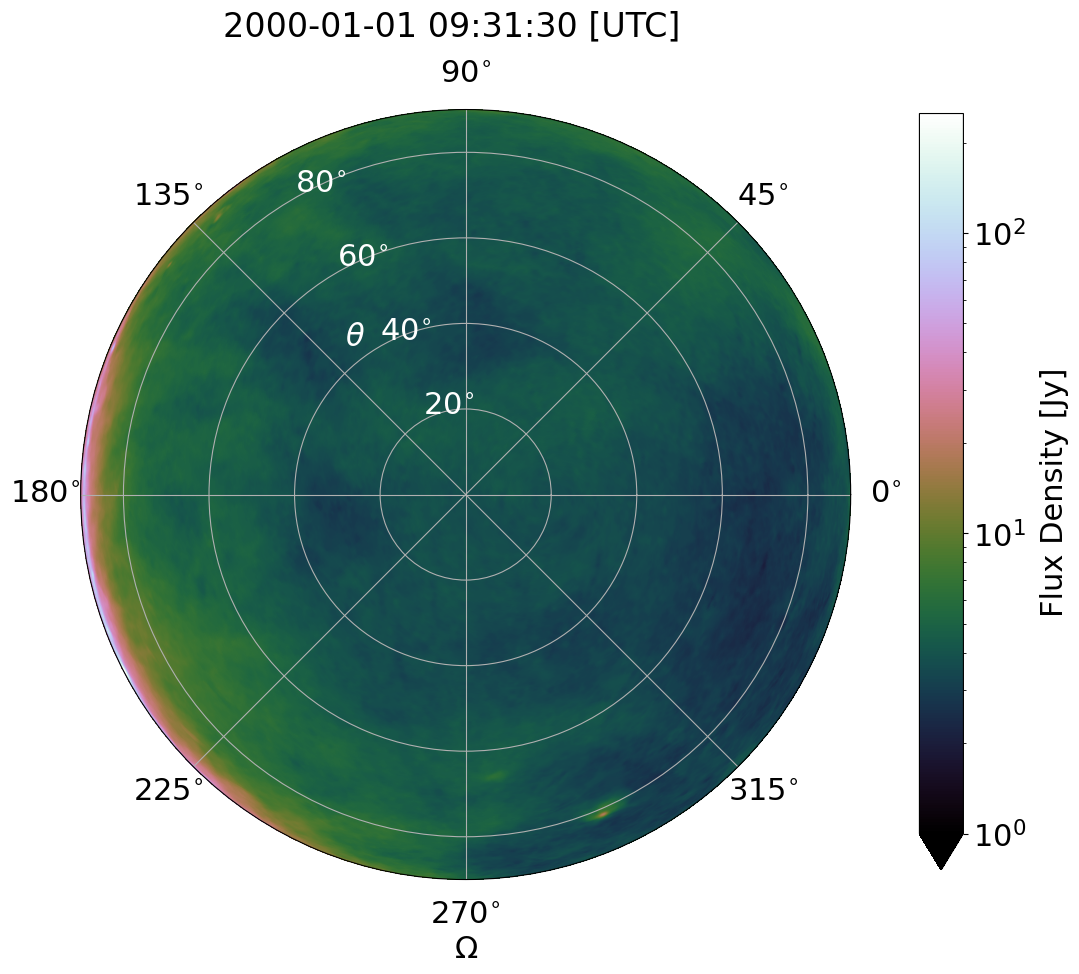}
    \end{subfigure}
    \begin{subfigure}{\linewidth}
    \includegraphics[width=1\linewidth]{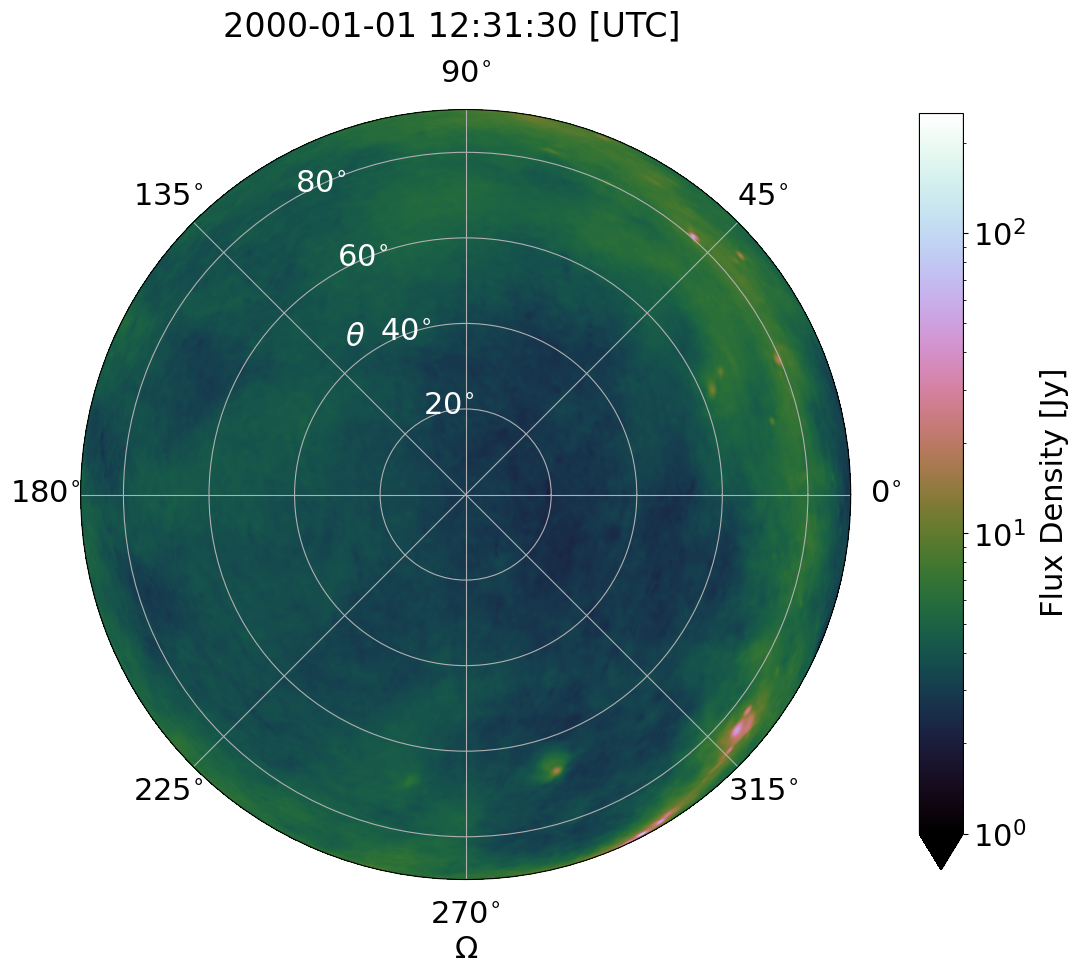}
    \end{subfigure}
    \caption{The composite sky model illustrating the source flux density as Dirac delta functions at 135~MHz, observed at two times: (top) 2000-01-01 09:31:30 [UTC] with the GC along the horizon and (bottom) 2000-01-01 12:31:30 [UTC] with the Galactic Centre below the horizon.}
    \label{fig:obstimes}
\end{figure}
For our simulations, we select two observation times. The first is at 12:31:30 UTC on January 1, 2000, when the Galactic Plane is far from the SKA-Low field of view and located along the horizon. The second is at 09:31:30 UTC on January 1, 2000, when the GC is below the horizon, minimising any leakage through the far-out sidelobes. The sky map at 135 MHz for both observation times is shown in Fig.~\ref{fig:obstimes} and a description of the sky model content is outlined in Section~\ref{sky model}.

\subsection{Telescope Model}

\subsubsection{Antenna geometry}
\begin{figure}
\centering
\includegraphics[trim=0.0cm 1.0cm 0.0cm 2.0cm,clip,width=7.5cm,height=7.5cm]{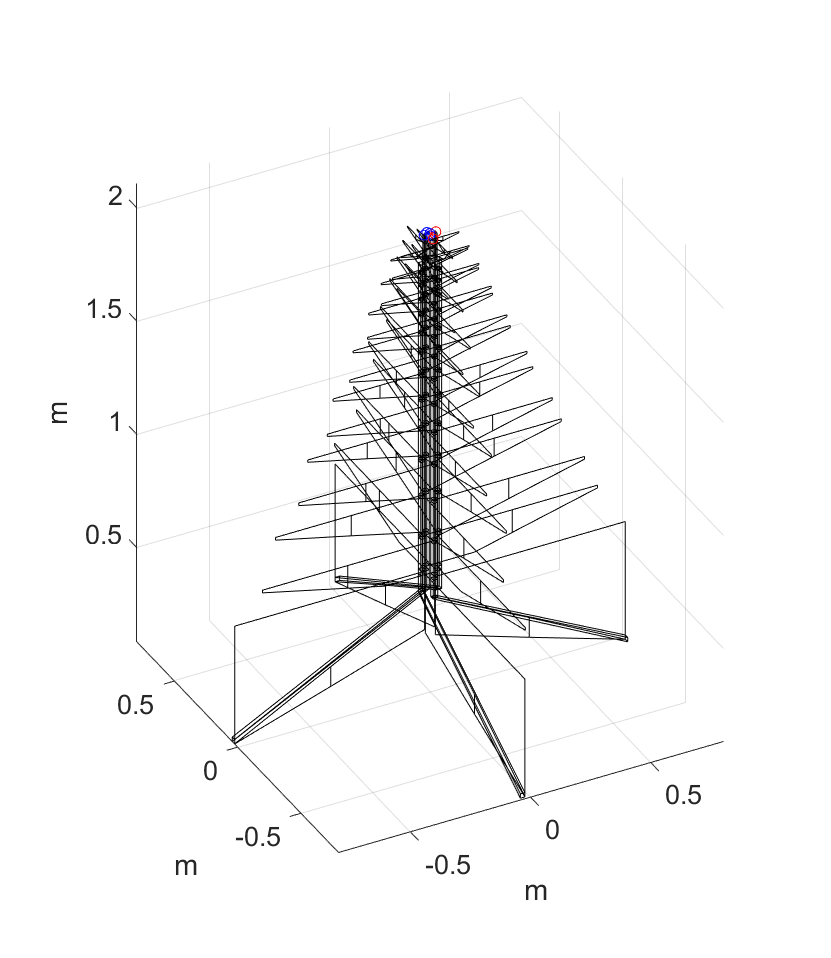}
\caption{Geometry of the SKALA4 \citep{de2018antenna}.}
\label{fig:SKALA4}
\end{figure}
The 4th version of the SKA element, illustrated in Fig.~\ref{fig:SKALA4}, is a dual-polarized log-periodic antenna featuring $16$ dipoles per arm (SKALA4, \citep{de2018antenna}). It operates over a $7:1$ frequency range, spanning from $50$~MHz to $350$~MHz. Each arm is differentially amplified using a low-noise amplifier (LNA) with an input impedance of $100$~Ohms.

\subsubsection{Station Layout}
\begin{figure*}
\centering
\includegraphics[width=1\linewidth]{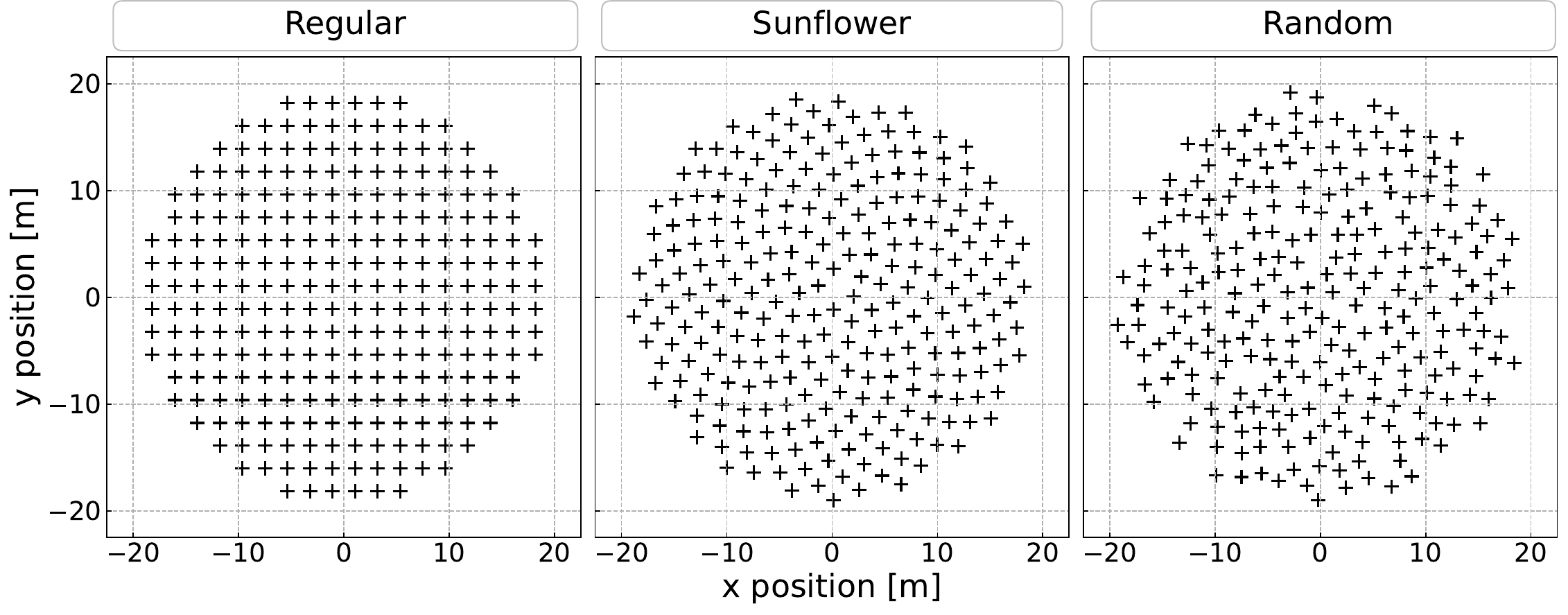}
\caption{The antenna positions within a SKA-Low station are shown for the ‘regular,’ ‘sunflower,’ and ‘random’ layouts.}
\label{fig:layouts}
\end{figure*}
Fig. \ref{fig:layouts} illustrates the three different layouts used in this numerical experiment. Each layout is arranged in a circular footprint with a radius of 19~m and a minimum distance of 1.7~m between the antennas.
\begin{itemize}
    \item \textbf{regular:} A regularly spaced grid of elements within the circular station footprint, each element positioned 2.14~m apart.
    \item \textbf{sunflower:} A sunflower-head array, with the position of each element given in \citet{VOGEL1979179} and evaluated for SKA-Low in \citet{9899989}.
    \item \textbf{random:} A random layout which results from random perturbation of the sunflower layout \citep{10501737}.
\end{itemize}

\subsubsection{Array Layout}
\begin{figure}
\centering
\includegraphics[width=1\linewidth]{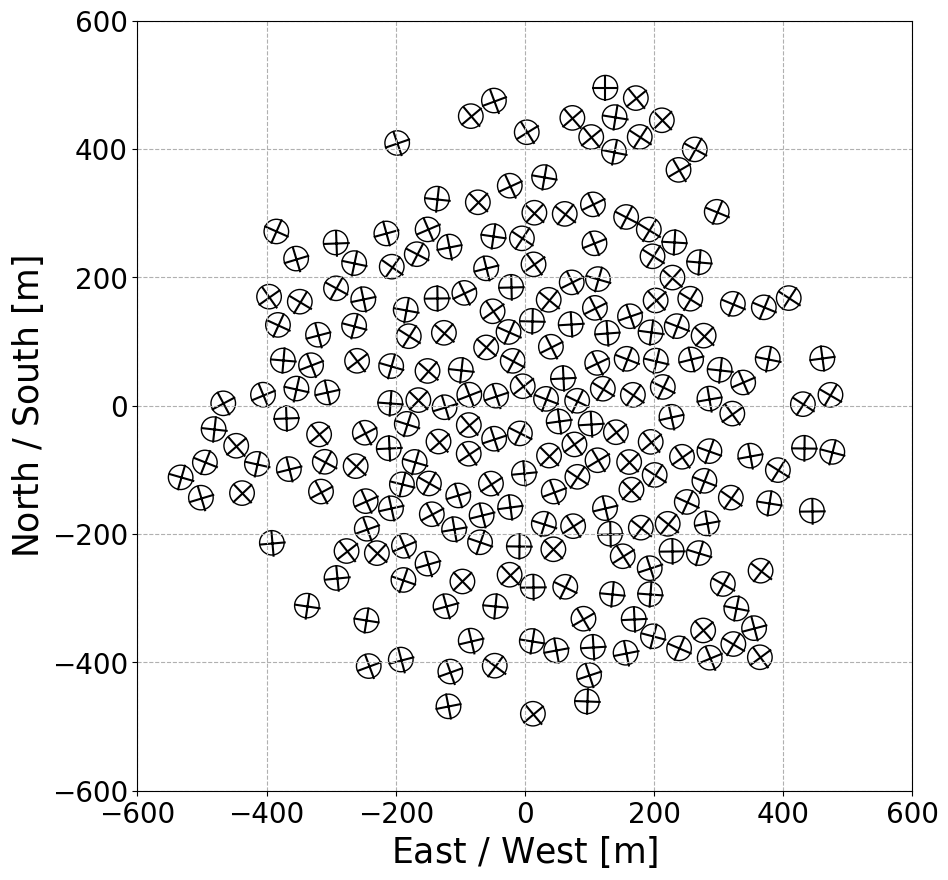}
\caption{The centres of the 224 stations in the SKA-Low core, each with a unique rotation applied. Baseline lengths range from approximately $40$~m to $1$~km.}
\label{fig:corelayout}
\end{figure}

The coordinates of the array centre and the projected spacing between these stations were obtained from Revision 3 of the SKA-Low configuration and illustrated in Fig.~\ref{fig:corelayout}. We exclude all the long baselines and limit our simulations to the 224 core stations of SKA-Low, to ensure that the currently available sky model meets the Nyquist resolution. Note also that, during our analysis, the baselines are irregularly binned based on a logarithmic length distribution specific to SKA-Low.

In the following, we consider the case where all stations in the core have an identical, un-rotated station layout. Additionally, we examine a scenario where each station (antenna orientation and position relative to the station centre) is rotated by a unique angle, as illustrated in Fig.~\ref{fig:corelayout}. The rotation angles are assigned randomly to each station, following a uniform distribution between $0$ and $2\pi$.

\subsection{Foreground Model}
\label{sky model}

To accurately model the localized and extended foreground emission accessible to the SKA-Low, a composite sky model is created by treating the extended emission as a set of unresolved point sources. The sky model components include:
\begin{itemize}
    \item A diffuse Galactic foreground emission from the desourced and destriped Haslam 408~MHz survey, with $\sim 5 \times 10 ^{7}$~pixels, equating to a 1.718\arcmin resolution \citep{remazeilles2015improved}.
    \item A collection of point sources\footnote{Note that the GLEAM catalogue contains A-team sources, such as Fornax A, which exhibit extended spatial scales that are challenging to model using point sources. For more accurate modelling, refer to \citet{line2020}.} from GaLactic and Extragalactic All-Sky MWA Survey (GLEAM), with $\sim 3\times 10 ^{5}$ sources \citep{Wayth_2015}.
\end{itemize}
At low frequencies, radio foreground observations may be approximated by a near-featureless power law across the frequency spectrum. The composite map can be frequency-scaled within OSKAR as needed using the equation:
\begin{equation}
    B(f, \hat{\mathbfit{s}}) = B(f_0, \hat{\mathbfit{s}})\left[\frac{f}{f_0}\right]^\beta.
\end{equation}
In this equation, $f_0$ denotes the reference frequency, and $\beta$ signifies the spectral index. For the Haslam 408~MHz survey, $\beta=-0.7$, whereas for the GLEAM survey, a distinct $\beta$ is provided for each source.

\section{Accelerated simulation tools}
\label{sec:pipeline}

We present the simulation pipeline used to generate the delay power spectrum for large interferometric arrays. Accurately modelling visibilities for telescopes with many antennas and stations poses computational challenges, as it requires full-wave simulations of each antenna's electromagnetic response, including MC effects, and the evaluation of APs over multiple directions and frequencies. This pipeline leverages two in-house tools: FAST for efficient electromagnetic simulations and the radio-telescope simulator OSKAR for parallelized interferometric calculations.

\begin{figure}
\centering
\begin{tikzpicture}[scale=\textwidth/17cm,samples=200]

\usetikzlibrary {arrows.meta}

\draw [draw=black] (0,0) rectangle (2,1.15);
\draw [draw=black] (3,0) rectangle (5,1.15);

\draw [black,-{Latex[length=3mm]},dashed] (1,0) -- (1,-1.5);
\draw [black,,-{Latex[length=3mm]},dashed] (4,0) -- (4,-1.5);

\draw [draw=black] (0,-3) rectangle (5,-1);

\draw [draw=black,dotted] (0.25,-2.75) rectangle (1.7,-1.5);
\draw [draw=black,dotted] (3.3,-2.75) rectangle (4.75,-1.5);

\draw [black,-{Latex[length=3mm]},dashed] (2.5,-3) -- (2.5,-4);
\draw [draw=black] (1.5,-5) rectangle (3.5,-4);

\draw [black,-{Latex[length=3mm]},dashed] (3.3,-2.5)  --  (1.7,-2.5);

\node[text width=3cm,align=center] at (4,0.6)
{FAST};
\node[text width=3cm,align=center] at (1,0.6)
{GLEAM\\+\\HASLAM};
\node[text width=2cm,align=center] at (2.5,-1.25)
{OSKAR};
\node[text width=3cm,align=center] at (1,-2.1)
{VIS SIM};
\node[text width=3cm,align=center] at (4,-2.1)
{HARP\\BEAM};
\node[text width=2.0cm,align=center] at (2.5,-4.5)
{Delay Power\\Spectrum};
\draw [black,-{Latex[length=3mm]},dashed] (2.5,-5) -- (2.5,-6);

\node[text width=3cm,align=center] at (4.5,-0.5)
{\textit{beam\\data}};
\node[text width=3cm,align=center] at (0.4,-0.5)
{$B(f_0,\hat{\mathbfit{s}})$};

\node[text width=2cm,align=center] at (2.5,-2.1)
{$\mathbf{AP}_p(f,\hat{\mathbfit{s}})$};

\node[text width=2cm,align=center] at (1.7,-3.5)
{$V_{pq}(f)$};
\node[text width=2cm,align=center] at (1.7,-5.5)
{$\tilde{V}_{av}(\tau,b)$};
\end{tikzpicture}
\caption{Flow-chart of the simulation pipeline}
\label{fig:pipeline}
\end{figure}
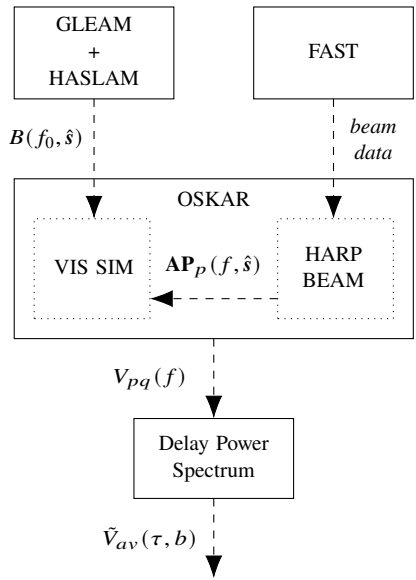

The pipeline, outlined in Fig.~\ref{fig:pipeline}, begins with electromagnetic simulations for the $2N_a$ EEPs in a frequency band $(f_{\min},f_{\max})$ at resolution $\Delta f$. Next, using sky catalogues and beam data from FAST, the OSKAR simulator computes the $N_s(N_s-1)/2$ visibilities for $N_s$ stations. During OSKAR visibility simulations, the beam computations are managed by the harp\_beam library (using FAST-generated beam data). Once all visibilities $V_{pq}(f)$ are generated, they are binned logarithmically by baseline length $b$ and averaged. We then apply a window function and perform the time-delay Fourier transform to obtain the time-delay visibilities $\tilde{V}_{av}(\tau,b)$. Computation times for both simulation tools will be based on the simulation parameters provided in Section~\ref{sec:simparameters}.

\subsection{FAST}
\label{FAST}
The recent advancements of FAST, \citep{gueuning2024FAST}, an accelerated direct solver based on the Method of Moments (MoM), have significantly reduced the simulation time required for the computation of all the EEPs of large, dense and irregular arrays of identical but complex antennas. In the case of SKA-Low, it provides all the EEPs for a frequency point in $\sim10$~minutes on a standard current laptop (see Table \ref{feko_fast_comparison} which details this benchmark). This represents a significant resource and time-saving improvement, as it operates approximately 25 times faster than FEKO's Multilevel Fast Multipole Method on a 128-core workstation. Additionally, re-simulation for different station layouts with FAST takes under a minute, allowing for the optimization of station layout and analysis of MC effects at a fine resolution. The electromagnetic solver has been validated against several other numerical solvers, including CST\footnote{See \url{https://www.3ds.com/products/simulia/cst-studio-suite}}, WIPL\footnote{See \url{https://wipl-d.com/}} \citep{bui2018fast}, IE3D\footnote{See \url{http://www.mentor.com/}} \citep{gueuning2019inhomogeneous} and FEKO\footnote{See \url{https://altair.com/feko}} \citep{gueuning2022inhomogeneous, gueuning2024FAST, 10643818}, showing agreement to within 2-3 digits of accuracy in the far-field pattern, provided the same mesh is used. Additionally, we have carefully validated our simulations against FEKO using smaller test-case stations to ensure that the chromaticity observed in our simulations results from MC and not from approximation errors related to the solver.

\begin{table}
\centering
\caption{Computation times to evaluate $512$ EEPs with FAST per frequency point.}
\begin{tabularx}{\columnwidth}{@{}l>{\centering\arraybackslash}X@{}}  
\toprule
\multicolumn{2}{c}{\bf Solver: Fast}\\
\midrule
\midrule
 station layout            & Random \\
 Machine                 & \mlinei{Intel Core i7-13800H, 2.50~GHz, 14 cores, 32.0~GB RAM} \\
 Pre-Computations   & 10.1~mins            \\
 Matrix Filling       & 0.23~mins            \\
 Solve Time        & 0.51~mins            \\
 Total Time         & 10.8~mins            \\
 Update Time per new layout   & 0.7~mins \\
 Peak Memory           & 1.3~GB             \\
\bottomrule
\end{tabularx}
\label{feko_fast_comparison}
\end{table}
\subsection{OSKAR}
OSKAR is a GPU-accelerated telescope simulator specifically designed for the SKA \citep{dulwich_fred_2020_3758491}, utilising the Radio Interferometric Measurement Equation (RIME). The RIME framework models the contributions of the complex interferometric visibility outlined in \eqref{Zernike} through a discrete sum across all visible sources in the sky whilst accounting for any potential instrumental effects including MC. To obtain the visibilities, OSKAR requires a star catalogue to describe all the source properties and a description of the position of each station, known as the sky model and telescope model respectively. For SKA-Low, this tool allows us to obtain the visibilities in just 4.3~seconds per frequency for $\sim 12.9\times 10^6$ sources (Table \ref{oskar_comparison} which details this benchmark). The harp\_beam library is integrated into OSKAR and efficiently calculates station beams directly from each antenna's current distribution computed from FAST. On an NVIDIA A100 GPU, it computes SKA-Low station beams for a 16,000-pixel grid in ~1 ms. Pre-computed APs save time for identical layouts, but differing layouts require re-computation. In such cases, OSKAR simulations show station beam evaluations dominate runtime, being nearly 1,000 times slower than visibility integrations.

\begin{table}
\centering
\caption{Computation times to simulate visibilities with OSKAR.}
\begin{tabularx}{\columnwidth}{@{}l>{\centering\arraybackslash}X@{}}  
\toprule
\multicolumn{2}{c}{\textbf{\bf OSKAR: Common Properties}} \\
\midrule
\midrule
Machine                   & \mlinei{2x AMD EPYC 7763 64-Core Processor 1.8~GHz, 1~TB RAM, \newline 4x NVIDIA A100-SXM-80~GB GPUs} \\
Source Catalogue         & Composite Sky Model: \# $\sim 12.9\times 10^6$ \\
Number of frequency channels & 300            \\
Number of time steps      & 1      \\
Station Layout            & Randomised-sunflower \\
Antenna Element           & Embedded SKALA4  \\
Number of Stations        & 224              \\ 
\bottomrule
\\
\end{tabularx}
\begin{tabularx}{\columnwidth}{@{}l>{\centering\arraybackslash}X>{\centering\arraybackslash}X@{}}  
\toprule
\textbf{Simulation}                  & \mlineii{\textbf{Station Beam Duplication}} & \mlineii{\textbf{Unique Station Rotation}} \\
\midrule
\midrule
Number of Unique Stations & 1        & 224  \\
Beam Evaluation Time [mins]   & 12.5     & 2268.8     \\
Visibility Calculation [mins] & 8.5  & 9.1   \\
Time Per Frequency [secs] & 4.3 & 455.6 \\
Total Simulation Time [mins] & 21.5  & 2277.9     \\ 
\bottomrule
\end{tabularx}
\label{oskar_comparison}
\end{table}

\section{Numerical Results}
\label{sec:numresults}
We present the numerical results obtained using our pipeline, as outlined in Section~\ref{sec:pipeline}, and the simulation parameters described in Section~\ref{sec:simparameters}.
First, we examine the variation in the BIR fall-off rate for three different station layouts, highlighting the impact of intra-station MC. Next, we explore a foreground avoidance scenario with the SKA telescope and discuss the size of the available detection window. Finally, we consider a foreground removal scenario in which the station beam is inaccurately modelled using either the AEP approximation, corrupted or interpolated EEPs. We then attempt to remove the foreground power using these approximate beam models.

\subsection{Beam Impulse Responses}
\label{subsec:BIRs}

\begin{figure}
\centering
\includegraphics[width=1\linewidth]{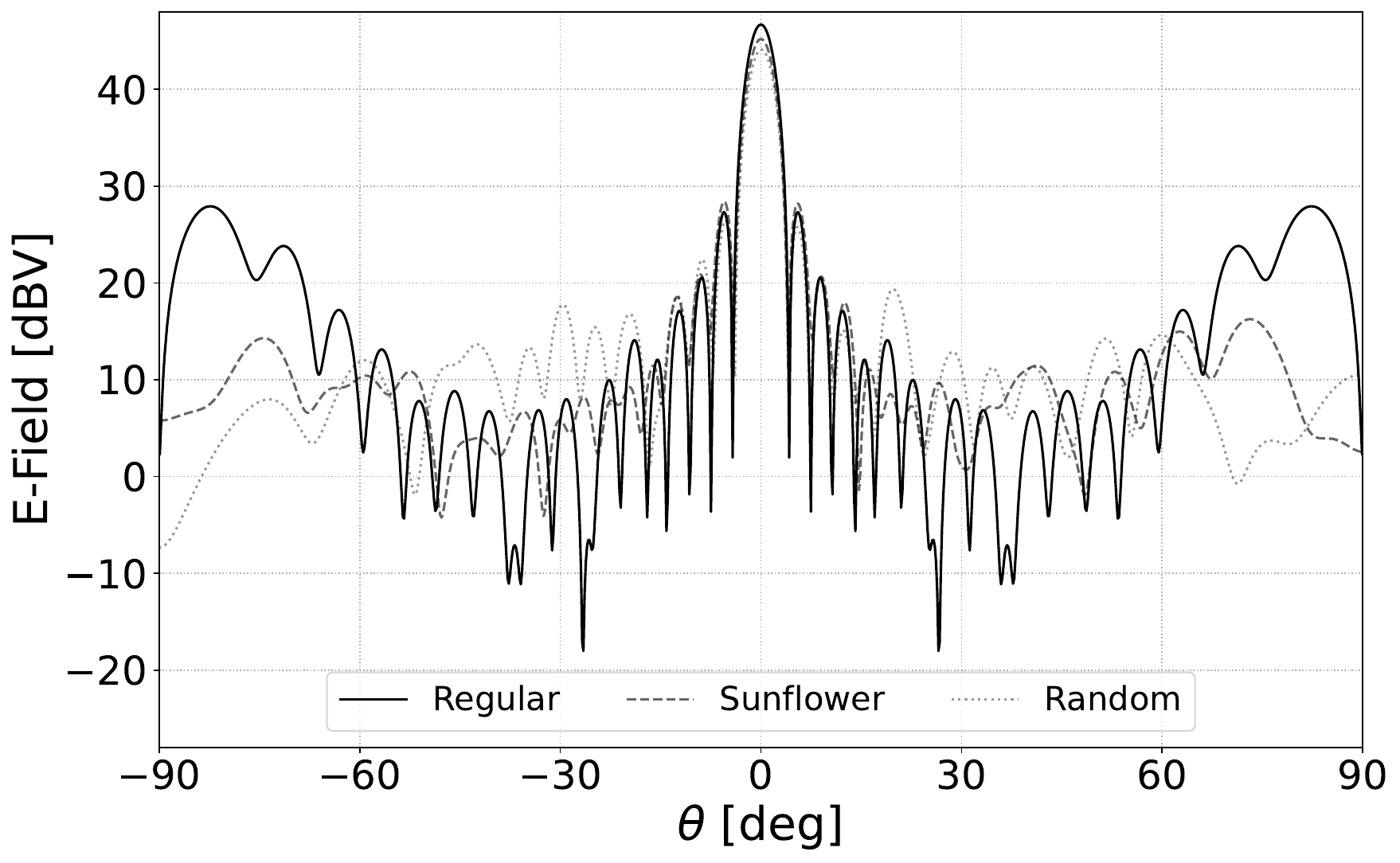}
\caption{H-plane cut of the SKA-Low station beams for the ‘regular’, ‘sunflower’, and ‘random’ layouts at 135~MHz.}
\label{fig:stationbeams}
\end{figure}

The station beams are plotted at $135$~MHz the band centre, in Fig.~\ref{fig:stationbeams} for the three different layouts illustrated in Fig.~\ref{fig:layouts}: regular, sunflower, and random. As can be expected \citep{6623095}, at this frequency, all three layouts exhibit a similar main beam value, around $45$~dBV, along with comparable levels for the first three sidelobes. The beam pattern for the regular layout exhibits large grating lobes (nearly $30$~dBV). The beam for the random layout shows higher secondary sidelobes compared to those of both the sunflower and regular layouts.

\begin{figure*}
\centering
\includegraphics[width=0.9\textwidth]{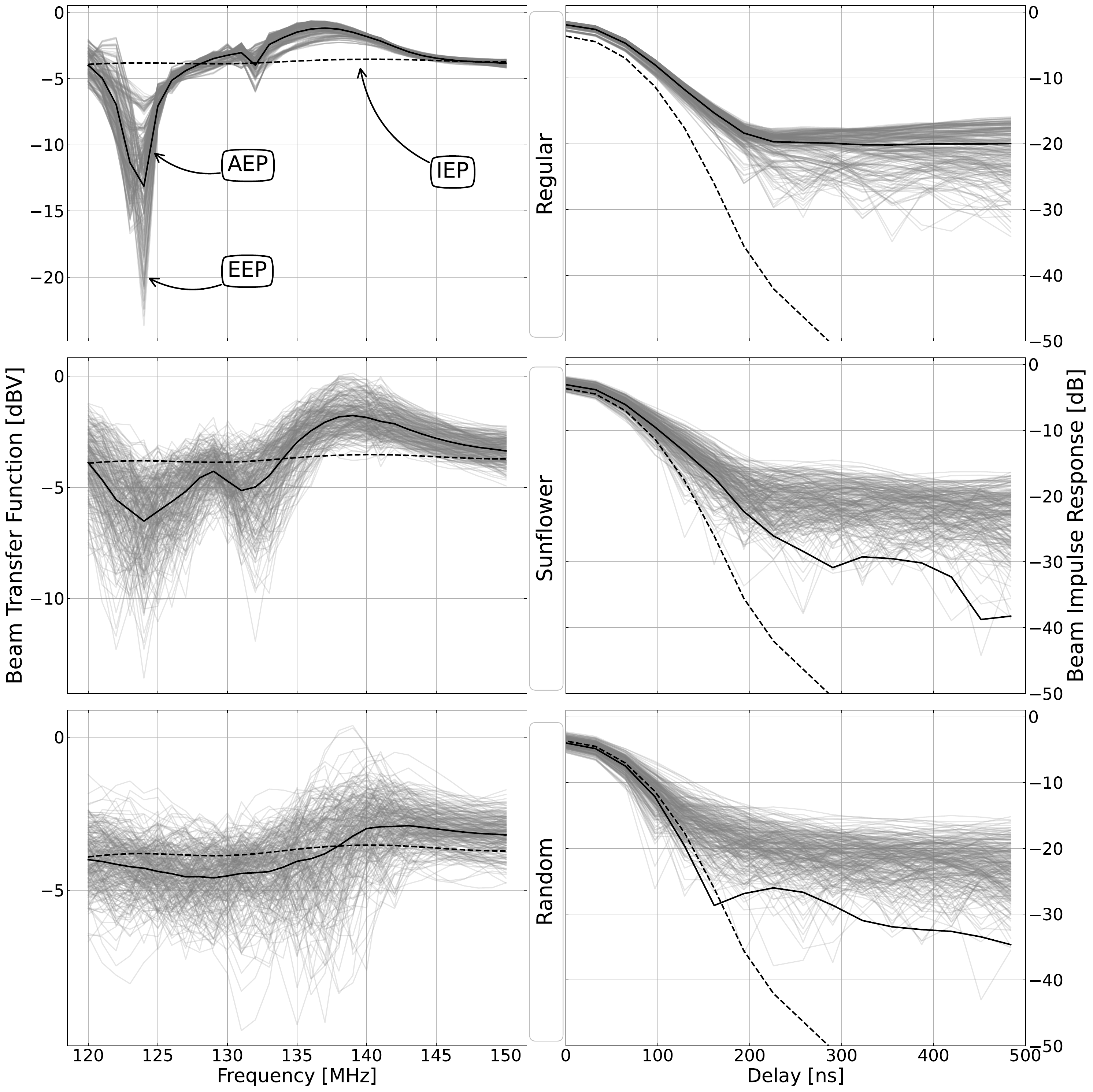}
\caption{The Beam Transfer Function (left) and Beam Impulse Response (right) for a SKA-Low station with regular (top), sunflower (middle), and random (bottom) layouts, evaluated at zenith. The results, shown for all EEPs (solid grey lines), the IEP (dashed black lines), and the AEP (solid black lines), are evaluated over 30 frequency channels ranging from 120~MHz to 150~MHz.}
\label{fig:beams}
\end{figure*}

Fig.~\ref{fig:beams} illustrates, in grey, all 256 individual EEPs for feed-H, evaluated at zenith and sampled at a 1~MHz rate, along with their corresponding BIRs. The solid black line represents the average embedded element pattern (AEP) across the elements, while the dashed line represents the isolated element pattern (IEP) of SKALA4. On the one hand, the IEP appears smooth across the frequency band, suggesting a nominal operation of the SKALA4 antenna. As a consequence, the BIR of the isolated antenna decays quickly reaching $-50$~dB at $300$~ns, with the main lobe width matching that of the window function $\tilde{W}$. On the other hand, prominent notches in the EEPs can be observed at $124$~MHz and $132$~MHz in the frequency response of both the regular and sunflower layouts. The drop at $124$~MHz results from destructive interference between the field radiated by currents on the active antenna and the field radiated by currents induced on the passive antennas due to MC. This effect occurs when the inter-element distance approaches one wavelength \citep{10702000}. The depth and width of this notch have been observed to increase with the array regularity, i.e., the redundancy in baseline length distribution \citep{10501737}. In the delay domain, these MC effects cause the first lobe of the BIRs of the EEPs to be wider for both the regular and sunflower layouts (extending to $200$~ns) compared to the first lobe width of the random layout (which extends to $150$~ns). The slower decay in the tail of the BIRs, at delays above $215$~ns, suggests the presence of sharp, small-scale spectral variations in the BTFs that vary at a quick rate (faster than 1 MHz). These variations appear to be less pronounced at the AEP level (or equivalently here in the AP) in the sunflower and random layouts. Specifically, the BIR of the AEP is around $-20$~dB for the regular layout, while it drops to $-30/40$~dB in the sunflower and random layouts. This attenuation is likely due to the irregular arrangement of antennas, which may help minimise MC effects at these scales.

\begin{figure*}
\centering
\includegraphics[width=1\textwidth]{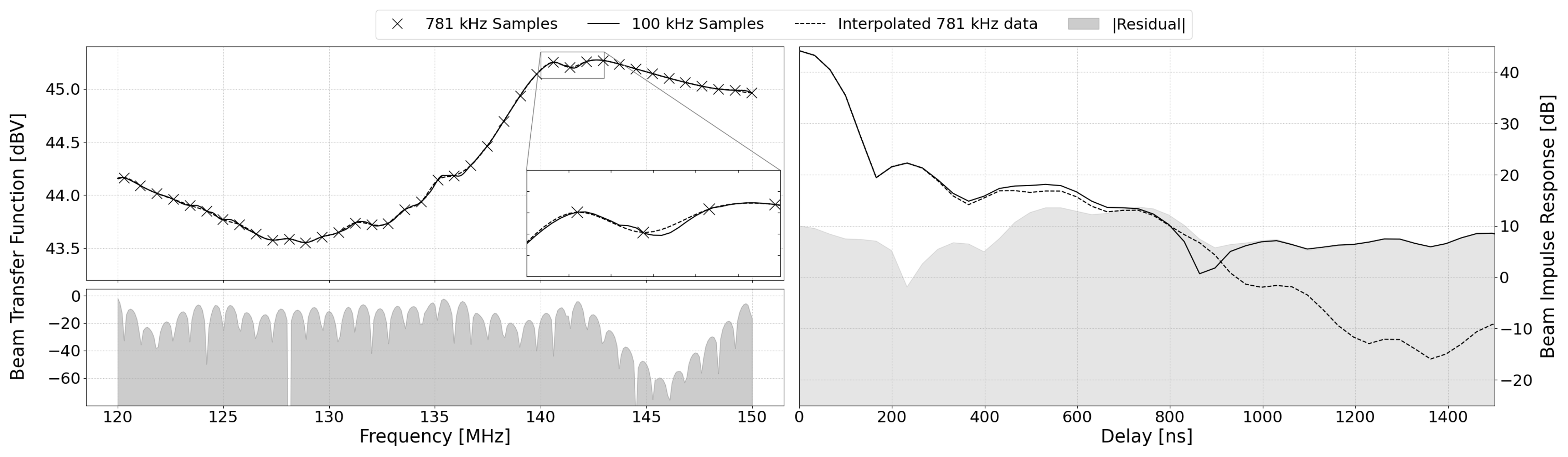}
\caption{The Beam Transfer Function (BTF, left) and Beam Impulse Response (BIR, right) simulated at 100~kHz (solid black line) and a cubic spline interpolated to 100~kHz from data sampled at 781~kHz (dashed black line) for a random layout SKA-Low station, evaluated at zenith. The BIR is truncated to 1500~ns. The BTF residual was initially calculated in volts and subsequently converted to dBV, whereas the BIR residual represents the Fourier transform of the power difference.}
\label{fig:stationbeambir}
\end{figure*}

We analyse now the BTF and BIR of the AP for the random layout, sampled at 100~kHz, and examine the impact of cubic interpolation from a coarse spectral grid, with steps of $\Delta f = {781}$~kHz, to match the expected SKA-Low station channel width \citep{trott2016spectral}. As shown in Fig.~\ref{fig:stationbeambir} (left), the zoomed region highlights rapid variations in the BTF that outpace the $781$~kHz rate, resulting in interpolation errors of up to $0$~dBV in the AP (roughly $2$ significant digits in the E-field pattern, respectively). In Fig.~\ref{fig:stationbeambir} (right), this results in time-delay residual errors around $10$~dB, leading to $3\%$ errors at small delays, and divergence from exact values at approximately $700$~ns. These interpolated EEPs will be applied in our pipeline for the foreground removal scenario in Section~\ref{subsec:removal} to evaluate the impact of interpolation errors on the delay power spectrum.

\subsection{Foreground avoidance}

Early studies suggest that the sensitivity of the SKA-Low telescope should be sufficient to detect the 21-cm power spectrum at scales defined by $k \in [0.1,0.33]$\,cMpc$^{-1}$ \citep{cohen2018charting, barkana2023anticipating}. However, this section illustrates, that MC effects may compromise these predictions.

To characterize foreground contamination, we define the delay spread, denoted as $\tau_{min}(b)$, also known as the foreground spill-over. This value represents the smallest delay at which the 21-cm signal power spectrum $\Tilde{\mathcal{P}}_{21}$ surpasses the foreground power spectrum $\Tilde{\mathcal{P}}_F$:
\begin{equation}
\label{min_excess_delay}
\tau_{\min} (b) = \min~[~\tau ~| ~\Tilde{\mathcal{P}}_F(b,\tau) < \Tilde{\mathcal{P}}_{21}(b,\tau)~].
\end{equation}
In our analysis, we calculate a fiducial 21-cm signal power spectrum amplitude \citep{lanman2020quantifying} of $\Tilde{\mathcal{P}}_{21} \approx 114^2~\mathrm{mK}^2$h$^{-3} \mathrm{Mpc}^3$ at the band's central redshift $z=9.52$\footnote{Each cosmological cube, generated at an integer redshift $z$, contains 21-cm brightness temperature samples $\mathcal{T}_{21}$, defined on $512^3$ pixel grid, with each pixel having a side length of 3~cMpc. To obtain the fiducial 21-cm signal power spectrum amplitude, we begin by calculating the wavevector $\mathbfit{k}$, given by Eq.~\eqref{kvec}. A three-dimensional discrete Fourier Transform is then computed on each simulation cube to obtain, $\mathcal{T}_{21}$ and the power spectrum calculated using, $\Tilde{\mathcal{P}}_{21}(k, z) = |\mathcal{T}_{21}|^2\times V_{\rm pix}^2/V_{\rm cube}$ averaged over 100 logarithmically-spaced $k$-bins in the range $[7\times10^{-3}, 2]$.}. This was derived using $1.5$~cGpc 21-cm brightness temperature cubes, generated with the semi-numerical code 21cmSPACE \citep{Visbal_2012, Fialkov_2014}\footnote{Information regarding the astrophysical parameters used to model the 21-cm cubes with 21cmSPACE can be found in Section~2.2 of \citet{o2024understanding}.}. 

\begin{figure}
\centering
\includegraphics[width=1\columnwidth]{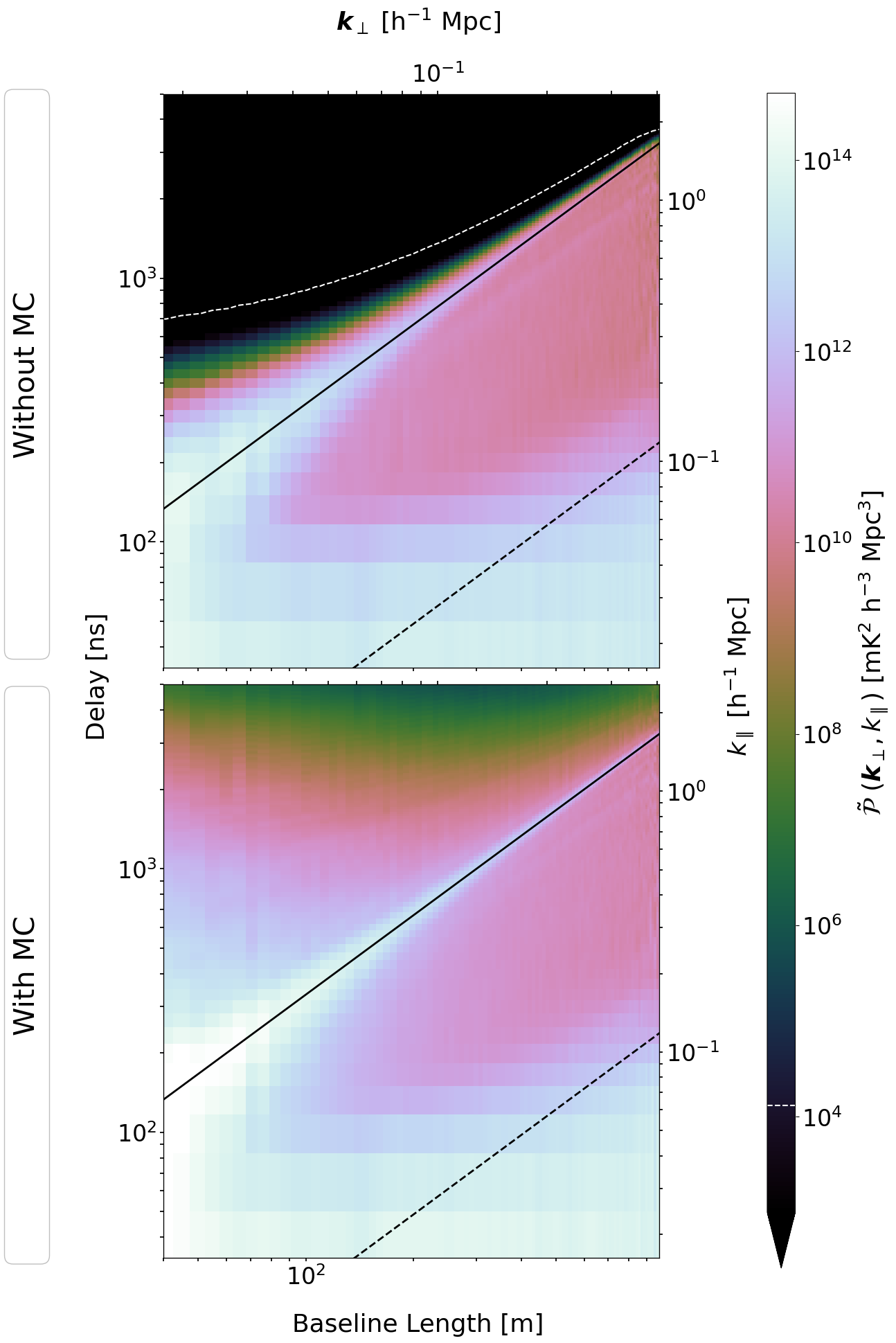}
\caption{The delay power spectrum at 09:31:30 UTC on January 1, 2000, without MC (top) and with MC (bottom), for the un-rotated random layout of the SKA-Low core across 300 channels from 120~MHz to 150~MHz. The solid black line represents the horizon limit, the dashed black line indicates the beam limit, and the dashed white line shows the foreground spill-over. Foreground contamination induced by MC completely masks the accessible Fourier modes of the detection window (EoR window).}
\label{fig:eep_vs_iep}
\end{figure}

Using the telescope m, Fig.~\ref{fig:eep_vs_iep} shows the delay power spectrum for an observation at 09:31:30 UTC on January 1, 2000, with the GC along the horizon. The simulation spans 300 channels from 120~MHz to 150~MHz, using an identical (un-rotated) random layout for all 224 stations of the SKA-Low core. The results are presented in two cases. In the first case, shown in the top plot (without MC), the AP is approximated using the IEP of SKALA4 multiplied by the Array Factor (AF). The AF is given by the equation:
\begin{equation}
\label{array_factor}
\textrm{AF}(f,\hat{\mathbfit{s}})  = \sum_{n=1}^{N_a}  e^{j k_0  \hat{\mathbfit{s}} \cdot \mathbfit{b}_{\textrm{ant},np} },
\end{equation}
where unit coefficients $c_{np}=1$ are assumed.
In the second case, shown in the bottom plot (with MC), the exact AP is used, which incorporates all EEPs as defined in \eqref{AP}. The dashed white line marks the delay spread, $\tau_{\min}$, outlining a 21-cm detection window shown as the upper left black triangular region in the top panel. In this region, the foreground signals are sufficiently attenuated, falling below the estimated level of the 21-cm signal. The solid black line indicates the horizon limit, defined by the maximum geometric delay (Eq.~\eqref{geo_delay}) for a source at the horizon.
A bright band of power, present across all baseline lengths and confined to low delays, is identified as the intrinsic foreground. Additionally, along the horizon limit, a bright limb is observed, caused by an apparent increase in flux density \citep{thyagarajan2015foregrounds}. The dashed black line indicates the beam limit, determined by the primary beamwidth of $4.2^{\circ}$ at the central frequency, as illustrated in Fig.~\ref{fig:stationbeams}. The slower decay rate of the BIR of EEPs, compared to that of the IEP, observed earlier in Fig.~\ref{fig:beams} translated into similar effects in the delay power spectrum plot shown in Fig.~\ref{fig:eep_vs_iep}. The delay spread for the smallest baseline ($\sim 40$~m) generated with the AF approximation extends to $700~$ns, while it extends beyond the maximum delay for the exact AP. The chromaticity introduced by MC significantly increases the delay spread across all baselines, causing the foreground emission to leak and completely occlude the 21-cm detection window. As validation of our absolute power spectrum levels, we compare the peak delay power spectrum without MC effects (top plot in Fig.~\ref{fig:eep_vs_iep}) against results from PAPER \citep{pober2013opening}, MWA \citep{beardsley2016first}, HERA \citep{Abdurashidova_2023} and SKA-Low simulation \citep{trott2016spectral}. The agreement confirms consistent diffuse emission delay spreads and comparable peak values (around $2 \ 10^{14}~\mathrm{mK}^2$h$^{-3} \mathrm{Mpc}^3$), despite differences in bandwidth and baseline binning.
We will next investigate how observation times, station layouts, and random station rotations influence the spill-over. 

\begin{figure*}
\centering
\includegraphics[width=1\textwidth]{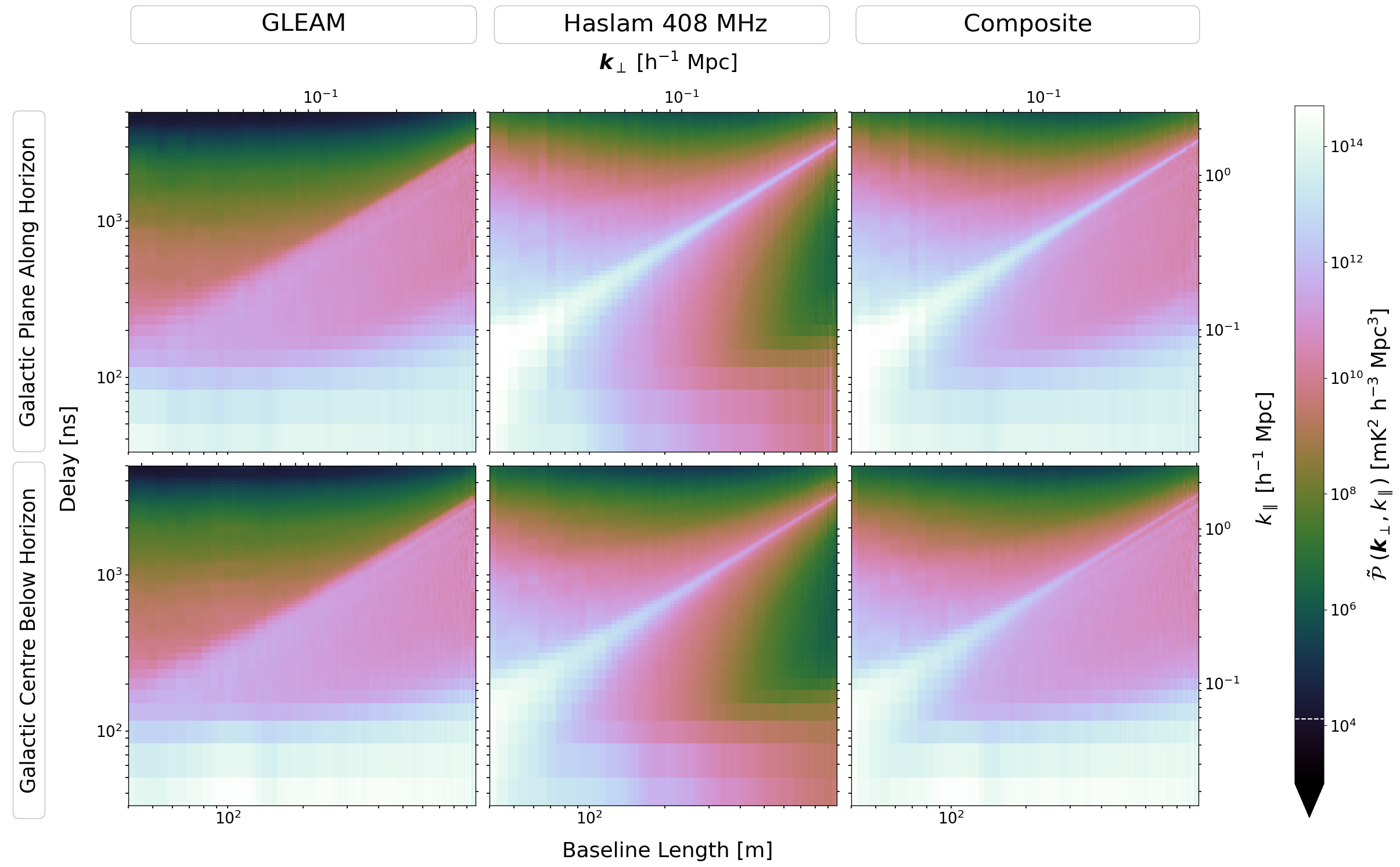}
\caption{The delay power spectrum for different sky map compositions (GLEAM, Haslam 408~MHz, and the composite sky map) is shown for two snapshots: at 09:31:30 UTC on January 1, 2000, when the Galactic plane is along the horizon, and at 12:31:30 UTC on January 1, 2000, when the Galactic Centre is below the horizon. Each observation was simulated with MC, using an un-rotated random station layout and the SKA-Low core across 300 channels from 120~MHz to~150 MHz.}
\label{fig:obs_source}
\end{figure*}

Using the same telescope model, Fig.~\ref{fig:obs_source} compares the delay power spectrum for two observations on January 1, 2000: at 09:31:30 UTC, when the Galactic plane is along the horizon, and at 12:31:30 UTC, when the GC is below the horizon. We decompose the delay power spectrum into separate contributions from point sources (using the GLEAM catalogue) and diffuse emission (using the HASLAM map). The results show that spill-over is larger for diffuse emission ($\sim10^{10}~\mathrm{mK}^2$h$^{-3} \mathrm{Mpc}^3$ at $2000$~ns) compared to point sources ($\sim 10^{6}~\mathrm{mK}^2$h$^{-3} \mathrm{Mpc}^3$ at $2000$~ns). This difference arises due to the increased chromaticity of the diffuse emission, resulting in an apparent broader delay spectrum $S(\tau)$. The diffuse emission contribution also tapers off as baseline length increases. Fig.~\ref{fig:obs_source} clearly shows that, at both observation times, diffuse emission is concentrated around the horizon line, while point source emission predominantly originates from emissions in the main beam. Close inspection of the left-hand panels (GLEAM) exhibits faint lines (an example is denoted by a black arrow) of increased power running parallel to the horizon line. Each line originates from an increase in sensitivity due to the fourth sidelobe of the random layout appearing in Fig.~\ref{fig:stationbeams}.
Furthermore, when the GC is below the horizon (bottom left plot), a notable increase in power is observed relative to when it is above the horizon (top left plot), extending to the beam limit. This increase is attributed to a higher source count and flux density within the primary beamwidth, as recorded by the GLEAM catalogue. In the central panels showing diffuse emission at both observation times, a decrease in diffuse power along the horizon is observed when the GC is below the horizon, compared to when it is at grazing angles, which explains the increased foreground spill-over when the GC is up.

\begin{figure*}
\centering
\includegraphics[width=1\textwidth]{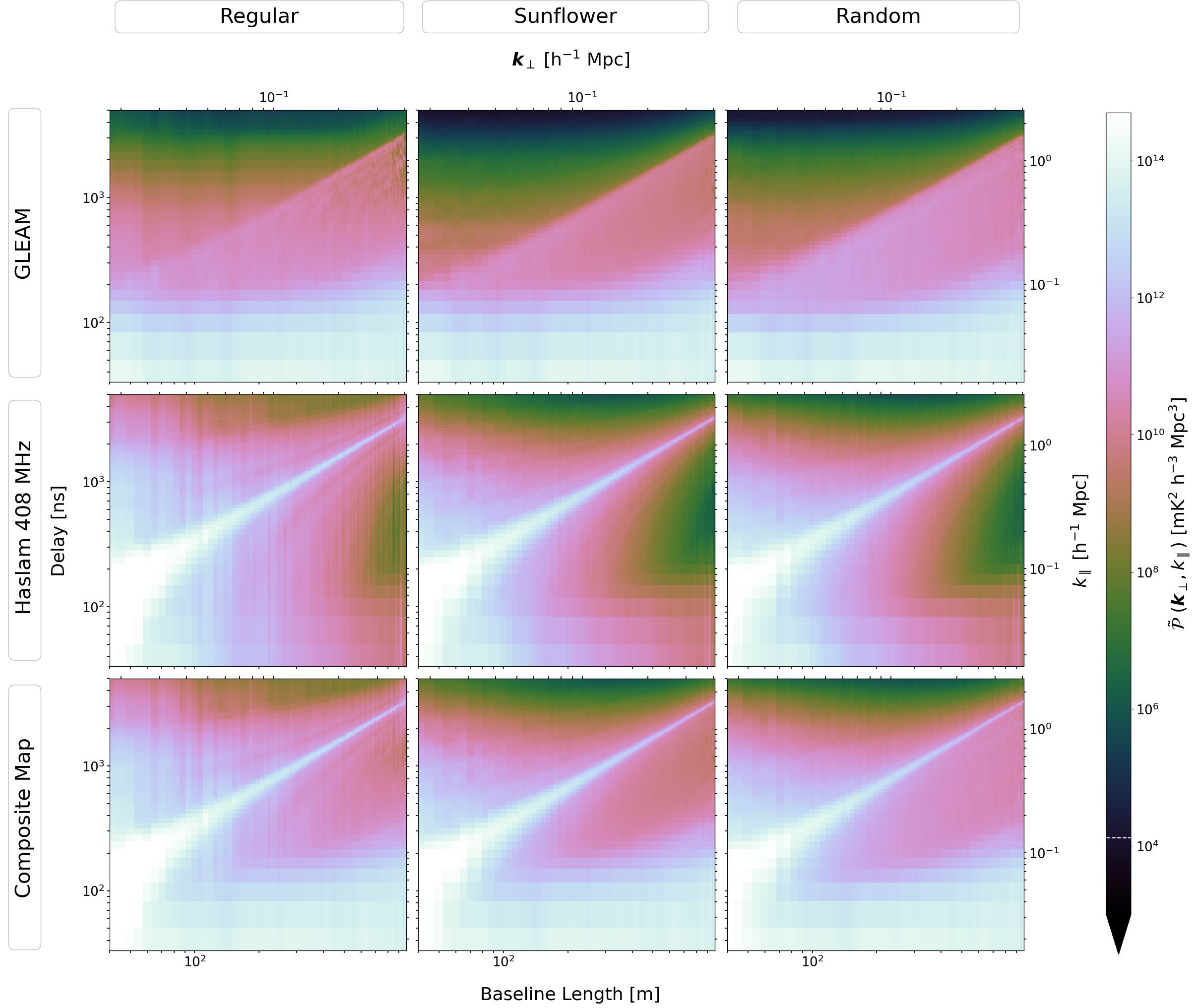}
\caption{Delay power spectrum across different sky map compositions (GLEAM, Haslam 408~MHz, and composite) and un-rotated station layouts (regular, sunflower, and random) including MC, for a snapshot at 09:31:30 UTC on January 1, 2000, with the SKA-Low core across 300 channels from 120~MHz to 150~MHz.}
\label{fig:layout_source}
\end{figure*}

In Fig.~\ref{fig:layout_source} we analyse how the power spectrum varies with identical station layout (regular, sunflower, and random, as shown in Fig.~\ref{fig:layouts}) for the first observation time with the GC above the horizon. Given that the main beam and first sidelobe levels are similar across layouts (see Fig.~\ref{fig:beams}), power levels remain consistent up to and slightly beyond the beam limit for both point source and diffuse emission. As the elevation angle increases to $20^\circ$, the larger sidelobes in the random layout compared to those in the sunflower or regular layouts result in increased point source power levels within the foreground wedge, between the horizon and beam limit, and with a more prominent sidelobe line corresponding to the contribution of bright point sources. The grating lobe of the regular layout appears as a brighter region near the horizon line.
Regarding MC-specific effects, in line with the discussions in Section~\ref{subsec:BIRs}, the regular and sunflower layouts exhibit a broader intrinsic foreground region, extending to approximately $200$~ns. In contrast, the random layout, with its narrower first lobe width, shows a response extending only to around $150$~ns. The delay spread is noticeably much larger for the regular layout for both point source and diffuse emission. The leakage of diffuse emission appears similar for the sunflower and random layouts. The spill-over of point source emission is however slightly lower for the sunflower layout (around $10^9~\mathrm{mK}^2$h$^{-3} \mathrm{Mpc}^3$ at $800$~ns) compared to the random layout (around $10^{10}~\mathrm{mK}^2$h$^{-3} \mathrm{Mpc}^3$ at $800$~ns). Overall, for all layouts, the entirety of the detection window (EoR window), up to $k_{\parallel} \sim 2~\mathrm{h}^{-1} \mathrm{Mpc}$ (see bottom row of Fig.~\ref{fig:layout_source}) is contaminated.

\begin{figure}
\centering
\includegraphics[width=1\columnwidth]{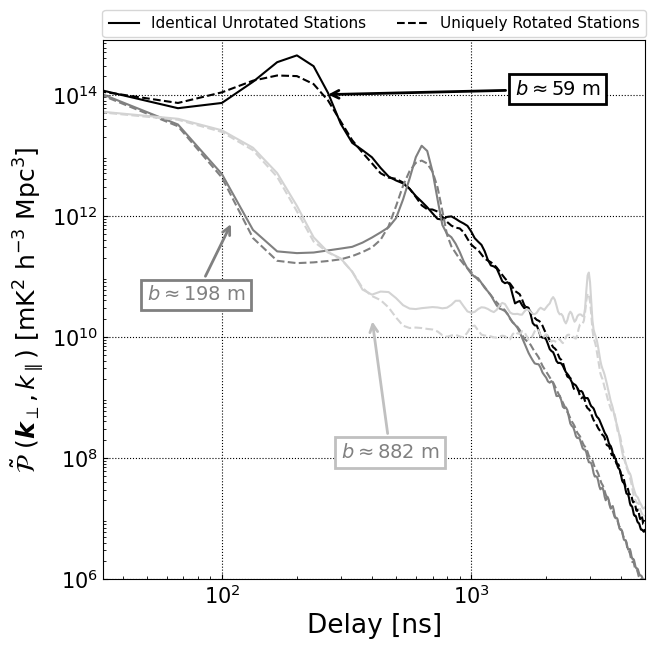}
\caption{A delay power spectrum slice for $b=[59,~198,~882]$~m was generated using 224 identical (un-rotated) stations and 224 uniquely rotated stations, both arranged in a random layout.}
\label{fig:rotation_slice}
\end{figure}

Each of the 224 SKA-Low core stations is now assigned a unique random rotation angle, uniformly distributed between $0^\circ$ and $360^\circ$ (see Fig.~\ref{fig:corelayout}). 
This rotation is expected to attenuate contributions from sources in the sidelobes for station layouts with more irregular azimuthal distributions. Practically, this means that each station beam becomes unique, requiring the computation of 224 individual station beams instead of just one, which significantly increases computation time in OSKAR, as shown in Table~\ref{oskar_comparison}. Fig.~\ref{fig:rotation_slice} compares the delay power spectrum slices for un-rotated and rotated random layouts at given baseline lengths $b = [59,~198,~882]$~m. While the size of the intrinsic foreground region remains unchanged, the rotated case shows a roughly twofold decrease in power beyond the beam limit (sidelobe region), particularly along the brightened horizon limb. However, leakage at longer delays remains unaffected.

\subsection{Foreground removal}
\label{subsec:removal}
The EoR program for the SKA-Low telescope \citep{koopmans2015cosmic} comprises three major experiments: power spectrum analysis, which uses visibilities averaged to around $100$~kHz and $5$~s; tomography, relying on imaging data formed from these visibilities; and the 21-cm Forest experiment, which employs high spectral resolution image cubes with finer spectral channels of $\sim 4$~kHz. Several studies \citep{chapman2016effect, 10.1093/mnras/stw1768, Li_2019} suggest that a combination of removal and avoidance approaches will likely be employed for 21-cm science with SKA-Low. As previously discussed, the foreground removal approach aims to eliminate the foreground contribution from measured visibilities, leaving uncontaminated 21-cm signals. Achieving accurate foreground removal requires a precise model of the instrumental response on the sky. To estimate foreground residuals, we conduct simple removal tests using approximate station beam models. The residuals in the visibilities are computed as:
\begin{equation}
\label{residual}
\mathrm{Res}~{V}_{pq}(f) = {V}_{pq}(f) - {V}_{pq,~\mathrm{app}}(f).
\end{equation}
where $\Tilde{V}_{pq,~\mathrm{app}}(f)$ represents the visibility computed with the approximate beam models. The power spectra associated with these residuals are then computed using \eqref{est_delaytransform} and \eqref{rel3}. In this study, we consider three approximations to the beam model: station beams corrupted by random noise, the AEP approximation, and the spectral interpolation of EEPs from coarser channels. These tests allow us to assess the impact of beam model inaccuracies on foreground removal and the resulting residuals in the delay power spectrum. In the following examples, we consider un-rotated station layouts.

\begin{figure*}
\centering
\includegraphics[width=1\textwidth]{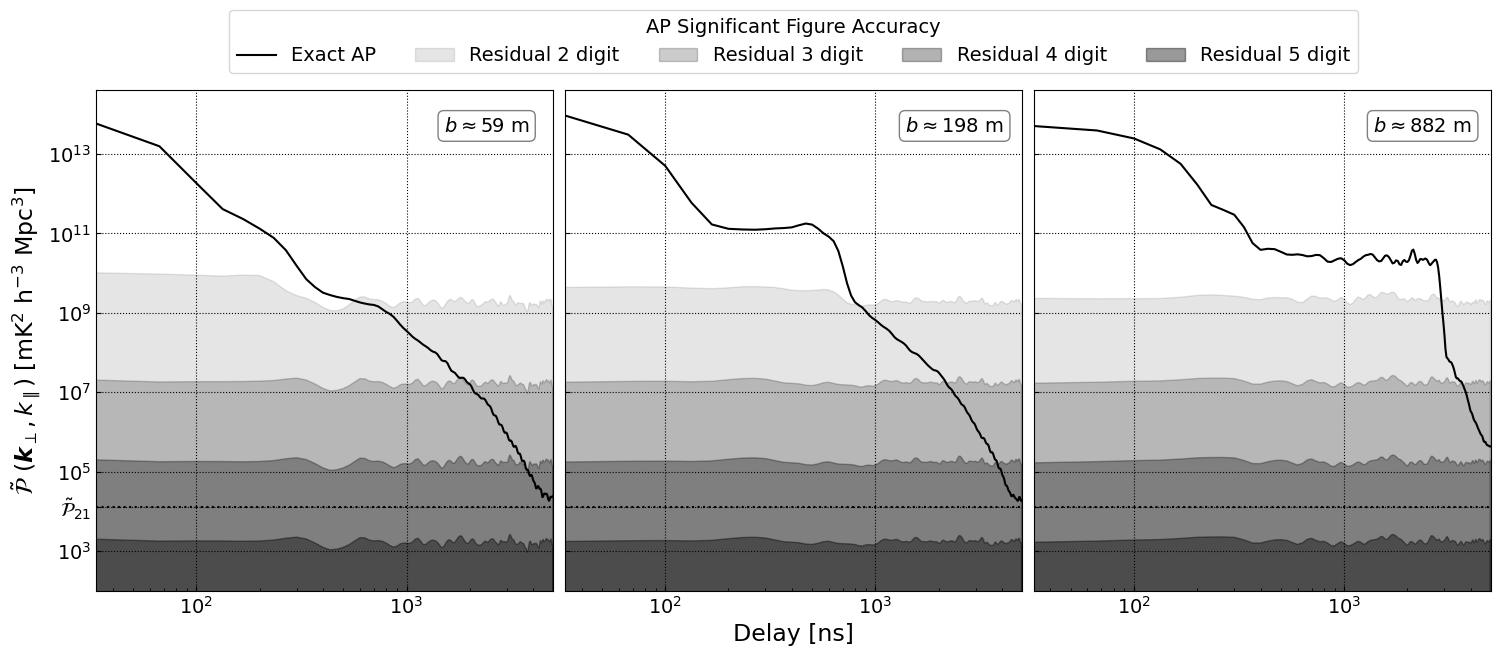}
\caption{Delay power spectrum slices of constant $b=[59,~198,~882]$~m, were simulated with an AP accuracy limited to 2, 3, 4, and 5 digits for an un-rotated random station layout. The residuals were computed with respect to the visibilities generated with the exact AP.}
\label{fig:noise}
\end{figure*}

The first test consists of adding Gaussian white noise to the APs as follows:
\begin{equation}
\label{ap_noise}
\mathbf{AP}_{p,~\mathrm{noise}}(f,\hat{\mathbfit{s}}) = \mathbf{AP}_p(f,\hat{\mathbfit{s}}) + \mathcal{N}_q(f,\hat{\mathbfit{s}}),
\end{equation}
where $\mathcal{N}_q$ represents complex-valued Gaussian noise with zero mean and variance $\sigma^2$. This noise is a multi-variable function of frequency $f$, the $N_{pix}$ pixel directions $\hat{\mathbfit{s}}$, and the station indices $p$. We assume four different values for the variances, $\sigma^2 = {10^{0}, 10^{-2}, 10^{-4}, 10^{-6}}$. For a signal coming from zenith, where the beam value is around $45$~dBV, this corresponds to signal-to-noise ratios $\textrm{SNR} = {45, 65, 85, 105}$~dB. Since $20$~dB corresponds to a single digit of accuracy in the E-field, the noise is approximately $2$, $3$, $4$ and $5$ digits below the received voltage for the respective SNR values. Figure~\ref{fig:noise} illustrates a selection of delay power spectrum residuals across baseline bins with lengths of $b = [59,~198,~882]$~m. Each additional significant figure of accuracy in the station model roughly corresponds to a reduction of two additional digits in the power spectrum residuals at small delays, with values around $(10^9, 10^7, 10^5, 10^3)$~mK$^2$h$^{-3}$Mpc$^3$ for the smallest baseline. Thus, to achieve effective foreground removal below our fiducial 21-cm signal, a 4-5-digit accuracy in the station beam model is required to keep the foreground residuals below the $114^2~\mathrm{mK}^2$h$^{-3} \mathrm{Mpc}^3$ level of the 21-cm signal. Since the AP is the sum of the EEPs (see \eqref{AP}), the noise variance on the EEPs that results in similar noise on the AP is $\sigma^2 / 256$. We estimate the required accuracy on the EEPs to be around 3-4 significant digits.

\begin{figure*}
\centering
\includegraphics[width=1\textwidth]{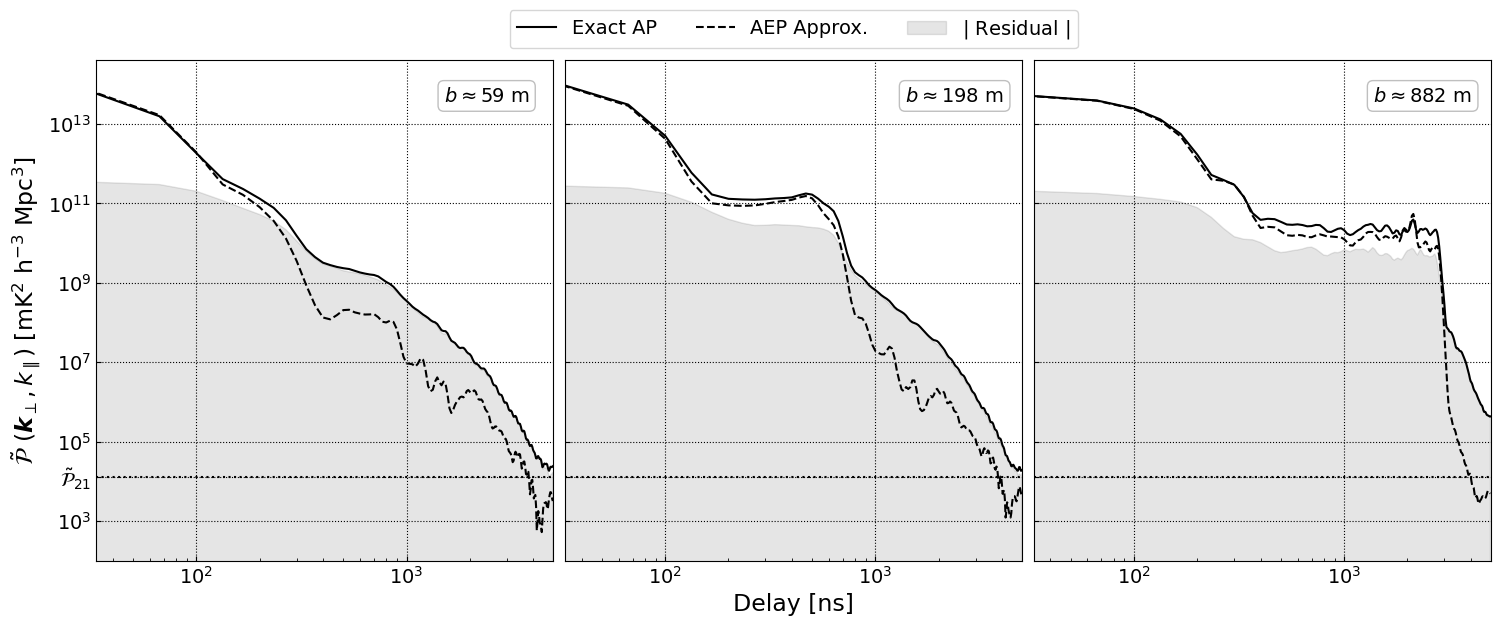}
\caption{Delay power spectrum slices of constant $b=[59,~198,~882]$~m for the exact AP, the AEP approximation and their absolute residual (Shaded grey region) for an un-rotated random station layout.}
\label{fig:AEP_approx}
\end{figure*}

Then, we analyse the residuals resulting from approximating the AP with the AEP multiplied by the AF \citep{wijnholds2019using} as follows:
\begin{equation}
\label{aep_approx}
\mathbf{AP}_p(f,\hat{\mathbfit{s}}) \approx \mathbf{EEP}_{av,p}(f,\hat{\mathbfit{s}}) \ AF(f,\hat{\mathbfit{s}}).
\end{equation}
Fig.~\ref{fig:AEP_approx} shows slices of the delay power spectrum for $b = [59,~198,~882]$~m, comparing the power spectrum using the exact AP, the AEP approximation, and the residual after attempting foreground removal in the visibilities. The AEP approximation is accurate within the intrinsic foreground region (up to $150$ ns) with residuals on the order of $1\%$ to $0.1\%$. 
The AEP approximation is off by $7$ orders of magnitude, preventing the residuals from reaching the level of our fiducial 21-cm signal. As shown in Fig.~\ref{fig:noise}, the residuals from the AEP approximation are $1$ order of magnitude higher than those obtained by using station beams only accurate to $2$ digits (assuming uncorrelated noise errors). Note that the tail of the estimated power spectrum decays faster than the true delay spectrum. This observation is consistent with the lower tail of the BIR associated with the AEP in Fig. \ref{fig:stationbeambir}.

\begin{figure*}
\centering
\includegraphics[width=1\textwidth]{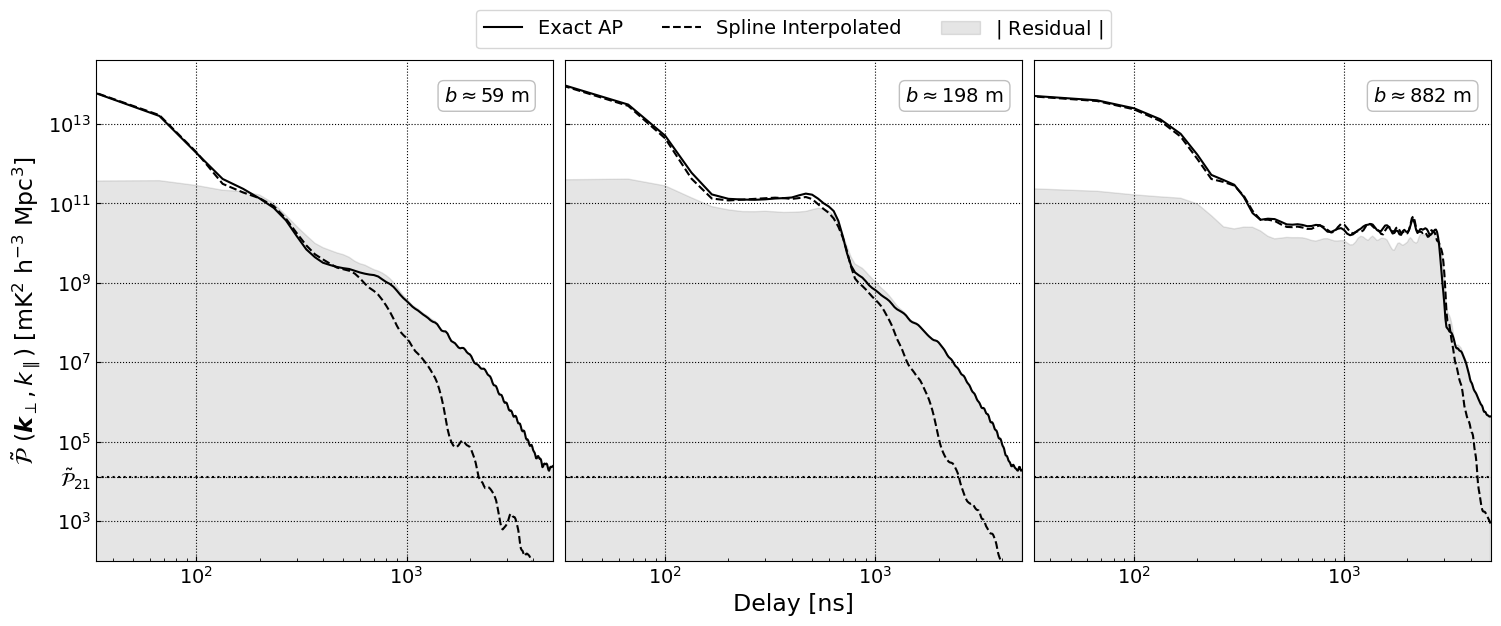}
\caption{Delay power spectrum slices of constant $b=[59,~198,~882]$~m using the exact AP, the AP with spline interpolated EEPs (upsampled from 781~kHz to 100~kHz) and their absolute residual (Shaded grey region) for an un-rotated random station layout.}
\label{fig:interp_approx}
\end{figure*}

In the final test, we examine the residuals resulting from interpolating the EEPs from a coarser spectral grid with a sampling rate of $\Delta f = 781$~kHz to the fine sample rate of $100$~kHz as discussed in Section~\ref{subsec:BIRs}. This coarse rate corresponds to the separation between the $384$ coarse channels of SKA-Low, covering the $300$~MHz bandwidth, which is formed at each station before applying beamforming and/or calibration weights \citep{comoretto2020signal}. Each coarse channel contains around $200$ fine $4$~kHz channels. The calibration strategy outlined in \cite{trott2016spectral} proposes using a low-order polynomial fit for the calibration coefficients across three adjacent coarse channels. Inspired by this work, we interpolate the EEPs, simulated at the centre of each coarse channel within the $120-150$~MHz band of interest, and then upsample these EEPs to a $100$~kHz resolution. The AP is then formed using \eqref{AP} from these interpolated EEPs, resulting in an approximate beam model. We have plotted the residuals for the three baseline lengths in Fig.~\ref{fig:interp_approx}. 
In agreement with the $30$~dB residual observed in the BIR of the AP (Fig.~\ref{fig:stationbeambir}), the residuals here range from $1\%$ to $5\%$. As shown in Fig. \ref{fig:interp_approx}, the coarse interpolation of the EEPs leads to an estimated delay power spectrum that decays faster than the true spectrum at long delays due to spectral smoothing. Considering the results of previous tests, another conclusion from this experiment is that reducing residuals by employing different EEP models for each antenna is effective only when these models are computed at high spectral resolutions.

\section{Conclusion}
\label{sec:conclusion}

This paper analyses the impact of Mutual Coupling (MC) between antennas on the time-delay power spectrum of the SKA-Low telescope. Electromagnetic simulations reveal significant and rapid spectral variations in the station beam directivity, driven by energy travelling and resonating between antennas due to MC. For 21-cm science, these MC effects cause substantial foreground leakage into the detection window, compromising foreground avoidance techniques. Highly accurate station beam models (within 4-5 orders of magnitude) are required to effectively remove foregrounds. Brute-force approximations, such as these based on randomising MC effects or on interpolating coarsely beam responses, will likely fall short. Although this study focuses on the direct evaluation of time-delay power spectrum, we expect image-based analyses, see for instance \citep{morales2019understanding}, to require similar beam model accuracy for image cleaning before power spectrum reconstruction. New calibration algorithms and measurement techniques will have to be developed to account for MC effects and enable precise beam modelling.

The analysis in this paper is made possible through the development of two key numerical tools:
\vspace{-0.1cm}
\begin{itemize}
\item The FAST electromagnetic solver: All EEPs were simulated across $300$ frequency points, completing in roughly two days on a standard laptop, with only a few additional hours needed for re-simulation in cases of new station layouts.
\item The OSKAR visibility simulator: The simulator processes the 24,976 visibilities in the SKA core across all frequencies in just $10$~mins using 4~A100~GPUs and including diffuse emissions and point sources across approximately 10 million sky pixels. The beam evaluation method embedded into OSKAR directly calculates station beams from the aperture current distribution, achieving evaluation times around $2$~seconds per station beam for the configuration above.
\end{itemize}
\vspace{-0.1cm}
Using these simulation tools, we fully integrated exact Array Patterns (APs) directly into visibility calculations, enabling precise tracking of MC effects in the time-delay power spectrum analysis. This yielded the following key findings regarding intra-station MC impacts:
\begin{enumerate}
\item Delay spread broadening: MC effects significantly extend the delay spread of the power spectrum. For EoR science, this can result in the complete obscuration of the detection window.
\item Layout-dependent MC effects: while offering beneficial beam redundancy, regular station layouts are more prone to resonant wave effects that lead to sudden power drops across the field of view. In contrast, randomized layouts reduce these issues, providing more stable beam behaviour and a narrower delay spread in the impulse response.
\item Residuals with AEP approximation: When approximating exact APs by using the AEP multiplied by the array factor, power spectrum residuals reach approximately $1\%$ within the foreground wedge, with peak values around $10^{11}~\mathrm{mK}^2$h$^{-3} \mathrm{Mpc}^3$. This is about $7$ orders of magnitude higher than the expected level of the 21-cm power spectrum ($\sim 10^{4}~\mathrm{mK}^2$h$^{-3} \mathrm{Mpc}^3$).
\item Residuals with corrupted station beams: Testing various levels of accuracy ($2$, $3$, $4$, and $5$ significant digits) resulted in residual power spectrum values of $10^9$, $10^7$, $10^5$, and $10^3~\mathrm{mK}^2$h$^{-3} \mathrm{Mpc}^3$, respectively. This demonstrates the need to model the station beams to $4-5$ significant digits. 
\item Residuals with interpolated EEPs: Interpolating the EEPs from a coarse ($781$~kHz) to a finer ($100$~kHz) spectral grid before forming the AP results in power spectrum residuals between $1\%$ and $5\%$, highlighting the need for high spectral resolution in EEP models to reduce interpolation errors.
\end{enumerate}

This work highlights the need to further develop station beam evaluation methods in processing pipelines, incorporate additional electromagnetic effects, and assess their impact across observational domains (baseline, delay, fringe rate, and image) to filter corrupted data and refine calibration algorithms. Future work should examine whether certain parts of the band are less affected by MC and extend these investigations to include image-based methodologies.

\section*{Acknowledgements}
The first two authors, Oscar S.D. O’Hara and Quentin Gueuning, contributed equally to this work and should be considered joint first authors.

The authors thank Dr. Karel Adamek for his assistance in parallelizing the HARP beam library, Dr. Vladislav Stolyarov for benchmarking it. We also thank Pr. Christophe Craeye for supporting the development of the fast electromagnetic solver, Dr. Maciej Syralek and Dr. Robert Laing for their support. We also thank the anonymous reviewers for their comments and suggestions, which have improved the quality of this paper.

This research was supported by ESA, NPL and the Science and Technology Facilities Council (STFC). OOH was supported by grant number G109464, ‘The Design of Highly Sensitive EM Sensors for Space Applications’, QG and FD by grant number ST/W00206X/1, EdLA by grant number ST/V004425/1 and DA and JC were supported by grant number ST/X00239X/1. AF is supported by a Royal Society University Research Fellowship \#180523. JD acknowledges support from the Boustany Foundation and Cambridge Commonwealth Trust in the form of an Isaac Newton Studentship.

\section*{Data Availability}
The data and software underlying this article will be shared on reasonable request to the corresponding authors.



\bibliographystyle{mnras}
\bibliography{main} 




\appendix
\section{The delay power spectrum}
\label{sec:DPS}
Here we provide a brief overview of the link between the power spectrum associated with a temperature field, the statistics we aim to measure, and the visibilities, which are the observable quantities. This relationship is straightforward, enabling the instrument to function almost as a direct probe of the signal power spectrum.
We begin with an estimation of the two-point correlation function, $\xi(\mathbfit{r})$ of a brightness temperature $\mathcal{T}(\mathbfit{r})$ as follows: 
\begin{equation}
    \xi(\mathbfit{r}') \approx \frac{1}{\mathcal{V}} \iiint_{\mathcal{V}} \mathcal{T}(\mathbfit{r})\ \mathcal{T}(\mathbfit{r} + \mathbfit{r}') \ \mathrm{d}r_x\mathrm{d}r_y\mathrm{d}r_z,
\end{equation}
where $\mathcal{V}$ is a comoving volume of interest \citep{hogg2000distancemeasurescosmology}.
The power spectrum $\mathcal{P}(\mathbfit{k})$ associated with this temperature field, expressed in [mK$^2$h$^{-3}$Mpc$^3$], corresponds to the Fourier transform of the correlation function $\xi(\mathbfit{r})$, which transforms comoving distances $\mathbfit{r}$ to wavevectors $\mathbfit{k}$ \citep{Parsons_2012, parsons2012per}:
\begin{equation}
\mathcal{P}(\mathbfit{k}) = \iiint_{-\infty}^{\infty} \xi(\mathbfit{r}) \ e^{-j \mathbfit{k} \cdot \mathbfit{r}} \ \mathrm{d}r_x\mathrm{d}r_y\mathrm{d}r_z.
\end{equation}
According to Parseval’s theorem, we have $\Tilde{\mathcal{P}}(\mathbfit{k}) = |\tilde{\mathcal{T}}(\mathbfit{k})|^2$ where $\tilde{\mathcal{T}}(\mathbfit{k})$ denotes the Fourier transform of the temperature field $\mathcal{T}(\mathbfit{r})$ \citep{Liu_2020}. 
The observed temperature $\mathcal{T}$ is associated with a specific point in co-moving coordinates and corresponds to an epoch defined by a particular redshift $z$.
We now express $\mathcal{T}$ as a function of local observational coordinates. Assuming the observed volume $\mathcal{V}$ is located at zenith and occupies a sufficiently small portion of the sky relative to the beamwidth, we can align the z-axis along the line of sight and apply the following transformation:
\begin{equation}
\label{rel1}
(l,m,f) = (r_x X(z),r_y X(z), r_zY(z)).
\end{equation}
This change of variable maps co-moving coordinates $(r_x,r_y,r_z)$ to observation coordinates $(l,m,f)$. Here, $X,Y$ depends on the redshift $z$ and are defined as \citet{lanman2020quantifying}:
\begin{equation}
X(z) \approx \chi(z) ,  \hspace{1cm} 
Y(z)  \approx  \frac{c_0 (1+z)^2}{H(z) f_{21}}
\end{equation}
where $\chi(z)$ is the co-moving distance, $H(z)$ is the Hubble parameter and $f_{21}$ is the rest-frame 21 cm frequency.
The relationship in \eqref{rel1} also establishes a mapping between the k-space and the observational baseline-delay domain: $(u,v,\tau) =  (k_x X(z)/\lambda_0,k_y X(z)/\lambda_0, k_zY(z))/(2\pi)$.
We decompose the wavevector $\mathbfit{k}$ into components parallel and perpendicular to the line of sight $\hat{\mathbfit{z}}$:
\begin{equation}
\mathbfit{k}_\perp = \dfrac{2\pi \mathbfit{u}_\lambda}{X(z)},  \hspace{1cm} k_{\parallel} = \dfrac{2\pi \tau}{Y(z)}.
\end{equation}
Using the notation $T(\hat{\mathbfit{s}},f) = \mathcal{T}(\mathbfit{r})$, we can express the Fourier transform with respect to the coordinates $(l,m,f)$ as follows:
\begin{equation}
\tilde{T}(\tau,\mathbfit{b}) = \iiint T(\hat{\mathbfit{s}},f) \ e^{-j 2\pi (\hat{\mathbfit{s}}  \cdot \mathbfit{b}_\lambda + \tau f)} \ \mathrm{d}l \mathrm{d}m \mathrm{d}f.
\label{rel2}
\end{equation}
Here, we thus have $\tilde{\mathcal{T}}(\mathbfit{k}) = \tilde{T}(\tau,\mathbfit{u})/(X^2Y)$ for purely planar baseline ($\mathbfit{b} = \mathbfit{u}$). The relationship in \eqref{rel2} is fundamental because $\tilde{T}(\tau,\mathbfit{b})$ can be nearly directly estimated from the time-delay transform of the instrument observable: the visibilities $V(f,\mathbfit{b})$. The main distinction is that the visibilities account for the instrument's response through the BTF, $A(\hat{\mathbfit{s}},f)$. When the width of the BIR, $\Tilde{A}(\hat{\mathbfit{s}},\tau )$ (in both angular $\hat{\mathbfit{s}}$ and time dimensions $\tau$), is significantly narrower than the temperature field $\Tilde{T}(\hat{\mathbfit{s}},\tau)$, we can approximate the power spectrum $\mathcal{P}$ as follows \citep{parsons2014new}:
\begin{equation}
\Tilde{\mathcal{P}}(\mathbfit{k}_\perp, k_\parallel) \simeq \dfrac{X(z)^2Y(z)}{\Omega_{A}} \ |\Tilde{V}(\tau,\mathbfit{b})|^{2},
\label{rel3}
\end{equation}
where $\Omega_{A}$ is a normalisation factor (a full derivation can be found in Appendix~B of \citealt{parsons2014new}) that accounts for the instrument response and is evaluated here as: 
\begin{equation}
\label{Omega_A}
\Omega_{A} = \int_{f_{\min}}^{f_{\max}} \iint |W(f) \ A(f,\hat{\mathbfit{s}})|^2 \ \frac{\mathrm{d}l}{n} \ \mathrm{d}m \ \mathrm{d}f
\end{equation}
The approximation \eqref{rel3} thus serves as an estimator of the power spectrum of a given signal of brightness temperature $T$.



\bsp	
\label{lastpage}
\end{document}